\documentclass[10pt,letterpaper]{article}

\usepackage{amsmath,amssymb}

\usepackage{graphicx}
\usepackage{booktabs}

\usepackage{changepage}
\usepackage{tabularx}

\usepackage{inputenc}

\usepackage{textcomp,marvosym}

\usepackage{cite}

\usepackage{nameref,hyperref}

\usepackage[right]{lineno}

\usepackage{tikz}

\usepackage{microtype}
\DisableLigatures[f]{encoding = *, family = * }
\usepackage[version=3]{mhchem}

\usepackage{array}

\usepackage{algorithm}
\usepackage{algpseudocode}

\usepackage{float}
\usepackage{timestamp}
\usepackage{xfrac}
\usepackage{mathtools}
\usepackage{arxiv}
\usepackage{chemformula}

\newcommand{\beginsupplement}{%
	\setcounter{table}{0}
	\renewcommand{\thetable}{S\arabic{table}}%
	\setcounter{figure}{0}
	\renewcommand{\thefigure}{S\arabic{figure}}%
	\setcounter{section}{0}
	\renewcommand{\thesection}{S\arabic{section}}%
}

\title{Autocorrelated measurement processes and inference for ordinary differential equation models of biological systems}
\newcommand\shorttitle{Autocorrelated measurement processes and inference for ODEs in biology}

\author{%
	Ben Lambert\\
	Department of Mathematics\\
	University of Exeter\\
	Exeter, UK\\
	\texttt{ben.c.lambert@gmail.com} \\
	\And
	Chon Lok Lei\\
	Institute of Translational Medicine\\
	Faculty of Health Sciences\\
	University of Macau\\ Macau, China\\
	\texttt{chonloklei@um.edu.mo}\\
	\And
	Martin Robinson\\
	Department of Computer Science\\
	University of Oxford\\
	\texttt{martin.robinson@cs.ox.ac.uk}\\
	\And
	Michael Clerx\\
	School of Mathematical Sciences\\
	University of Nottingham\\
	\texttt{michael.clerx@nottingham.ac.uk}
	\And
	Richard Creswell\\
	Department of Computer Science\\
	University of Oxford\\
	\texttt{richard.creswell@hertford.ox.ac.uk}
	\And
	Sanmitra Ghosh\\
	MRC Biostatistics Unit\\
	University of Cambridge\\
	\texttt{sanmitra.ghosh@mrc-bsu.cam.ac.uk}
	\And
	Simon Tavener\\
	Department of Mathematics\\
	Colorado State University\\
	\texttt{tavener@math.colostate.edu}
	\And
	David J. Gavaghan\\
	Department of Computer Science\\
	University of Oxford\\
	\texttt{david.gavaghan@cs.ox.ac.uk}
}

\begin{document}

\maketitle

\section{Abstract}
Ordinary differential equation models are used to describe dynamic processes across biology. To perform likelihood-based parameter inference on these models, it is necessary to specify a statistical process representing the contribution of factors not explicitly included in the mathematical model. For this, independent Gaussian noise is commonly chosen, with its use so widespread that researchers typically provide no explicit justification for this choice. This noise model assumes `random' latent factors affect the system in ephemeral fashion resulting in unsystematic deviation of observables from their modelled counterparts. However, like the deterministically modelled parts of a system, these latent factors can have persistent effects on observables. Here, we use experimental data from dynamical systems drawn from cardiac physiology and electrochemistry to demonstrate that highly persistent differences between observations and modelled quantities can occur. Considering the case when persistent noise arises due only to measurement imperfections, we use the Fisher information matrix to quantify how uncertainty in parameter estimates is artificially reduced when erroneously assuming independent noise. We present a workflow to diagnose persistent noise from model fits and describe how to remodel accounting for correlated errors.

\section{Introduction}
Ordinary differential equation (ODE) models are used throughout biology, typically to describe dynamic processes. Amidst a huge range of applications, ODEs are used to describe the transmission dynamics of infectious diseases \cite{anderson1992infectious}; they can represent the dynamics of enzyme-catalysed reactions \cite{murray2007mathematical}; and can explain the formation of action potentials in neurons \cite{hodgkin1952quantitative}. In ODE models, the evolution of a system depends only on its current state and a set of input parameters, which determine how individual components of the system interact. The parameters of ODE models in biological systems are typically not directly measurable and must be inferred from data. In this paper, we consider the assumptions underpinning inference of parameters from biological data.

A typical ODE model for modelling a dynamic process may be written:
\begin{equation} \label{eq:ivp}
\begin{aligned}
\frac{dx}{dt} &= h(t, x, \theta), \qquad t \in (0, T], \\
         x(0; \theta) &= x_0,    
\end{aligned}
\end{equation}
where $x(t; \theta) \in \mathbb{R}^n$ is the state of the system, $\theta\in \mathbb{R}^m$ are the parameters of the system, $t$ denotes time, $h(t, x, \theta)$ can be a function of time, state and parameters, and $x_0 \in \mathbb{R}^n$ is the initial state.

We suppose that an ODE model is proposed to explain a dataset: $\{\tilde y(t_i)\}_{i=1}^N$, where $\tilde y(t_i)\in\mathbb{R}^{l}$ and $l\leq n$. By fitting the model to these data, an analyst hopes to recover estimates of the parameters, $\theta$, which incorporate uncertainty. ODE models typically do not explain all variation within a dataset because they are approximations of the underlying processes, meant only to capture the most dominate characteristics of variation. Particularly in biology, the measurement of the system itself is also imperfect: measurement apparatus has a finite resolution and may provide indirect measures of the quantity of interest, and human errors may also contribute noise to observations. Because of these factors, a random error process is hypothesised to connect noisy observations with the ODE solution. This may be written:
\begin{equation}\label{eq:g_measurement}
    y(t_i) = g(x(t_i)) + \epsilon(t_i),
\end{equation}
where $g:\mathbb{R}^n \rightarrow \mathbb{R}^l$ allows a measured quantity to be a function of the ODE solution. In eq. \eqref{eq:g_measurement},  $\epsilon(t_i)$ is a random variable that represents both the effects of model misspecification and measurement noise.

The canonical assumption for the error terms is that they represent independent and identically distributed (IID) draws from a normal distribution \cite{ashyraliyev2009systems,mendes1998non,gabor2015robust,vanlier2013parameter,villaverde2019benchmarking,girolami2008bayesian}: $\epsilon(t_i) \stackrel{\text{IID}}{\sim}\mathcal{N}(0,\sigma)$, where $\sigma>0$ characterises the width of this distribution. The IID normality assumption is so widespread that it is typically stated without justification.

The normality assumption may be justified on the basis of a central limit theorem if it is thought that a series of independent or weakly dependent random variables -- representing different characteristics of measurement and misspecification processes – contribute additively to the overall errors; it may also be reasonable since the normal distribution emerges from a disparate range of processes representing measurement imperfections \cite[chapter~7]{jaynes2003probability}. But, if there is strong correlation between these constituent parts, then a distribution with heavier tails, such as a Student-t distribution or a Huber distribution is more appropriate \cite{maier2017robust}.

An IID normal distribution can also be justified by invoking the principle of maximum entropy \cite{jaynes2003probability,simoen2013prediction}. This principle roughly states that a probability distribution representing the outcomes of a process of interest should be chosen to include as little possible information about a process subject to known constraints. If only the mean and variance of the outcomes of a process are known, and there is thought to be zero correlation between errors, then it can be shown that an IID normal distribution is the probability distribution that makes the fewest additional assumptions \cite[chapter~7]{jaynes2003probability}. But it is unclear how applicable this is to the error distribution for ODEs, since we typically know only that the mean of the error distribution is zero, and our empirical examples indicate that the independence assumption may be an unreasonable null hypothesis. In particular, if there is thought to be autocorrelation in the noise, then a multivariate normal over the errors is the distribution with maximum entropy.

\begin{figure}[!ht]
	\centerline{\includegraphics[width=0.4\textwidth]{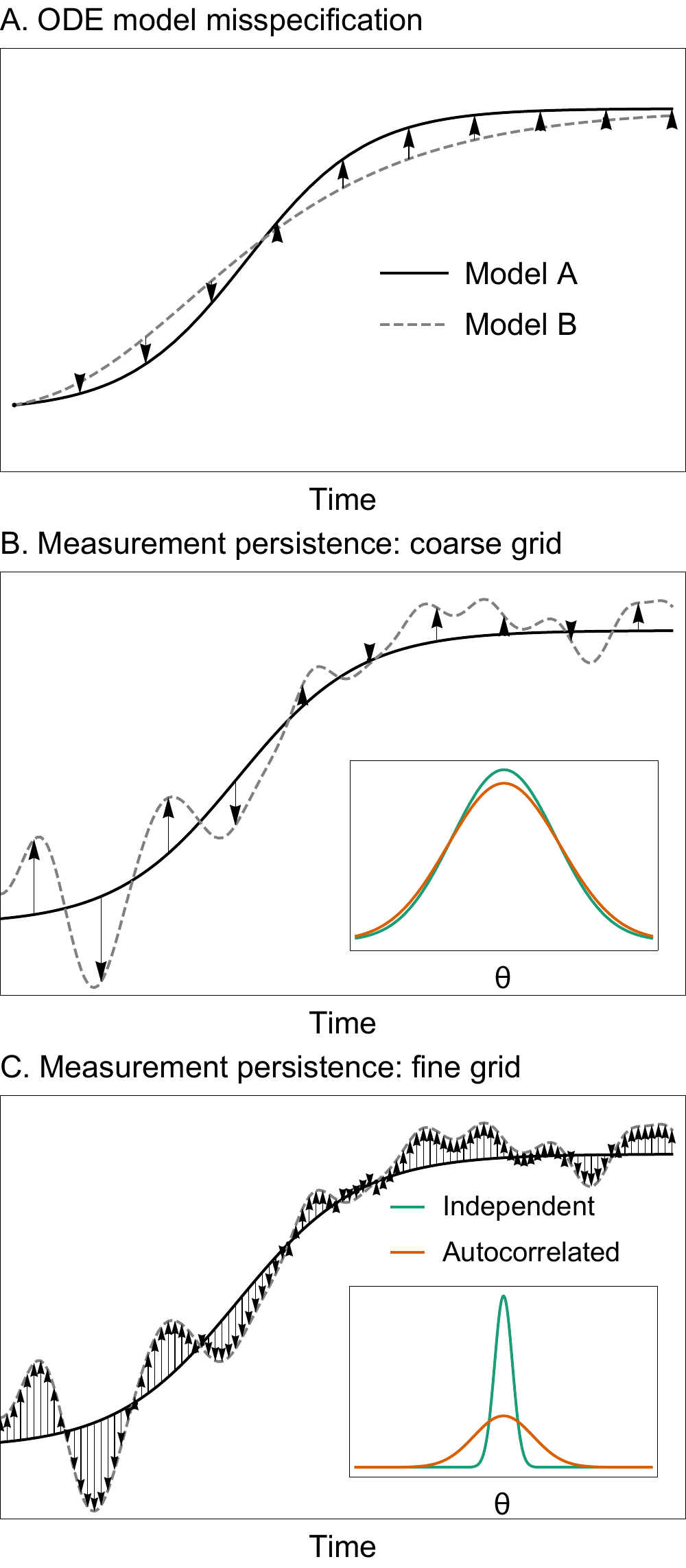}}
	\caption{\textbf{Causes of autocorrelated noise.} Panel A. shows how using a logistic model when, in fact, a Gompertz model is correct, results in autocorrelated noise; Panels B. and C. show how an imperfect measurement process can lead to different characteristic residual noise processes: in panel B., the measurements are taken using a coarse grid; in panel C., the measurements are taken using a fine grid. In both, residuals are depicted by black arrows. The inset plots show representative posterior distributions under different assumptions about the measurement process.}
	\label{fig:causes}
\end{figure}

There are two general causes of autocorrelation in the errors: misspecification of the model and poor temporal resolution of the measurement process \cite{simoen2013prediction}. In Fig. \ref{fig:causes}A, we illustrate how misspecifying an ODE model can lead to autocorrelated errors. This figure shows the outputs of two dynamic models as solid (model A) and dashed (model B) lines. We suppose that there is no measurement noise and that the data (arrow tips) is generated by model A. In attempting to fit these data, suppose model B is mistakenly chosen, and its best fitting line is as shown in this panel. There are manifold ways in which a model can be misspecified: the assumed functional form governing interactions between variables can be incorrect; important variables can be left out of the model entirely; a deterministic model may be used when a stochastic one is more appropriate; and so on. In this example, any of these issues could conceivably result in the differences between model A and model B, and, by choosing model B, this misspecification results in residuals (shown as arrows) exhibiting positive autocorrelation.

There is a huge literature devoted to accounting for model misspecification during inference (see, for example, \cite{kennedy2001bayesian,brynjarsdottir2014learning,lyddon2018nonparametric,lei2020considering}), and this remains an active area of research. In this paper, however, we focus only on the impact of assumptions around measurement noise, since, as we demonstrate, these can have dramatic effects on inference even in the absence of model misspecification. To exemplify how measurement process imperfections can lead to autocorrelation, suppose again that model A is the true model of nature, and that we (correctly) use it as part of our model of the data generating process. Also, suppose that the measuring apparatus is imperfect, producing noisy observations that may differ from the true underlying state, and has finite temporal resolution meaning it struggles to capture changes in output over shorter time scales. In Figs. \ref{fig:causes}B\&C, we show the model solutions (solid lines) and the values that would be measured if using a very fine temporal gridding (dashed lines). A consequence of this smooth measurement process is that the more observations per unit time are taken, the greater the degree of autocorrelation in residuals. In Fig. \ref{fig:causes}B, we show coarse observations of the system of interest as indicated by the horizontal positioning of the vertical arrows. In this case, since observations are sufficiently separated in time, there is relatively low persistence in residuals. In Fig. \ref{fig:causes}C, we take more observations of the same process, which produces positively autocorrelated residuals.

Intuitively, when the measurement process is positively autocorrelated, each observation conveys less information about the system than when the observations are uncorrelated. So misrepresenting an autocorrelated error process with one assuming independence can lead to overly confident parameter estimates. This is a well-known result in regression modelling \cite{wooldridge2015introductory}, and, since fitting ODE models to data is just nonlinear regression, these results should also apply to inference for these model types. We show this in the inset panels in Figs. \ref{fig:causes}B\&C: here, the orange lines show (illustrative) posterior distributions resultant from modelling the measurement process correctly; the green lines show the distributions when modelling the measurements assuming independence amongst them. In Fig. \ref{fig:causes}B, where the measurements are widely spaced, there is little difference in the recovered posteriors due to the limited autocorrelation. In Fig. \ref{fig:causes}C, failure to account for autocorrelation results in a posterior with too little variance.

We originally became interested in the impact of measurement autocorrelation on parameter estimation when attempting inference for a model of an electrochemistry experiment. Specifically, we noticed that the estimates obtained were unrealistically precise when assuming an IID normal error model, and the errors were autocorrelated. This led us to consider how this phenomena might be more generally applicable and whether there were guiding principles of how the degree of overconfidence depends on measurement autocorrelation. Thus, in this paper, we explore how measurement autocorrelation affects the precision of estimates. Previous work, in the context of modelling physical systems, has derived straightforward expressions for parameter uncertainty for a dataset consisting only of two time points with an accordingly simple error model \cite{simoen2013prediction}. Here, we consider a much more general setting where the models are nonlinear ODEs, which is typical in biological systems analysis, and the measurement process can be any one of a wide class of stochastic processes. We also account for the bias in the estimates of the standard deviation of the noise when fitting a model assuming IID Gaussian errors, which is important to ensure correct estimates of the degree of overconfidence. Using simulated data from ODE models, we demonstrate the validity of our analytical results. Using experimental data from cardiac physiology and electrochemistry, we show that highly persistent differences between observations and modelled quantities can occur. Whilst only illustrative, these results hint that overconfidence in parameter estimates may not be uncommon. In addition, we provide a workflow for diagnosing and accounting for autocorrelated errors when fitting an ODE model to data.

\section{Effect of autocorrelated noise on parameter estimate uncertainty}\label{sec:methods_autocorrelated}
In this section, we use mathematical analysis to evaluate the effect on parameter estimates of not accounting for autocorrelation when present. To do so, we first calculate ``\textit{true}'' parameter uncertainties obtained when specifying a persistent error model. We then calculate ``\textit{false}'' uncertainties obtained when assuming independent errors. To derive these quantities, we calculate the Fisher Information matrix (FIM) in both circumstances. This analysis shows that uncertainty in parameter estimates is understated when (falsely) assuming independent errors, with the degree of overconfidence increasing along with the persistence of the true errors. We call the ratio of true parameter estimate variance to that estimated assuming independent errors the ``\textit{variance inflation ratio}'' (VIR).

In \S\ref{sec:methods_constant}, we estimate the VIR for the mean parameter of a simple model with constant mean, when the actual error process is persistent and described by an autoregressive order-one (AR(1)) process. Calculating the VIR for the constant mean model is straightforward but provides a useful guide when examining more realistic cases. In \S\ref{sec:methods_nonliear_ode}, we consider a nonlinear ODE model with AR(1) measurement noise. In \S\ref{sec:ma1}, we explore the consequences of more ephemeral autocorrelations by calculating the VIR for the constant mean model with moving average order-one (MA(1)) errors. Realistic noise processes are likely, in fact, to be combinations of persistent and transient correlated noise, and in \S\ref{sec:arma}, we give formulae for computation of VIRs in this, more general, case.

\subsection{Constant mean model}\label{sec:methods_constant}
In what follows, we assume a time series framework where, at time $t$, observed data, $x(t)$, differs from its true constant value, $\mu$, by an additive random component,
\begin{equation}\label{eq:simple_system}
x(t) = \mu + \epsilon(t),
\end{equation}
where $\epsilon(t)$ is a zero-mean error random process such that ${\mathbb{E}[x(t)]=\mu}$.

There are a number of ways that measurement errors may be autocorrelated, and, in this paper, we consider a range. To begin, we consider AR(1) errors, in which there are persistent deviations between the observations and the true values of a process. This could occur, for instance, if a measurement apparatus responds slowly to changes in a system, meaning observations taken closer together are likely to be correlated due to measurement imperfections. An AR(1) process can be represented mathematically by:
\begin{equation}\label{eq:ar1}
\epsilon(t) = \rho \epsilon(t-1) + \nu(t),
\end{equation}
where $\nu(t)\stackrel{\text{IID}}{\sim}\mathcal{N}(0,\sigma)$, and $-1<\rho<1$ characterises the degree of autocorrelation: positive values indicating positive autocorrelation; and similarly so for negative values.

We first derive the \textit{true} (asymptotic) variance of the maximum likelihood estimator of $\mu$ when assuming an AR(1) error process in accordance with the true generating process. To do so, we use the log-likelihood to determine the diagonal element of the FIM corresponding to $\mu$ when we assume $\rho$ is known. To write down the log-likelihood, we require an expression for $\nu(t)$ in terms of the observables and parameters of the system, which can be obtained by multiplying $x(t-1)$ given by eq. (\ref{eq:simple_system}) by $\rho$ and subtracting it from $x(t)$, resulting in: ${\nu(t) = x(t) - \rho x(t-1) - \mu(1 - \rho)}$. Since $\nu(t)$ is distributed as an independent Gaussian, the log-likelihood of the model for a sample of observations ${x(t):\forall t\in[0,1,2,...,T]}$ is given by,
\begin{equation}\label{eq:simple_log_likelihood}
\mathcal{L} = -\frac{T}{2} \text{log }2\pi - \frac{T}{2} \text{log } \sigma^2 - \frac{1}{2\sigma^2}\sum_{t=1}^{T}(x(t) - \rho x(t-1) - \mu(1 - \rho))^2.
\end{equation}
Where, for simplicity, we have assumed that $\nu(0)=0$ is fixed and known -- \S\ref{sec:model_fitting} describes an alternative likelihood that does not make this assumption.

\begin{figure}[ht]
	\centerline{\includegraphics[width=1\textwidth]{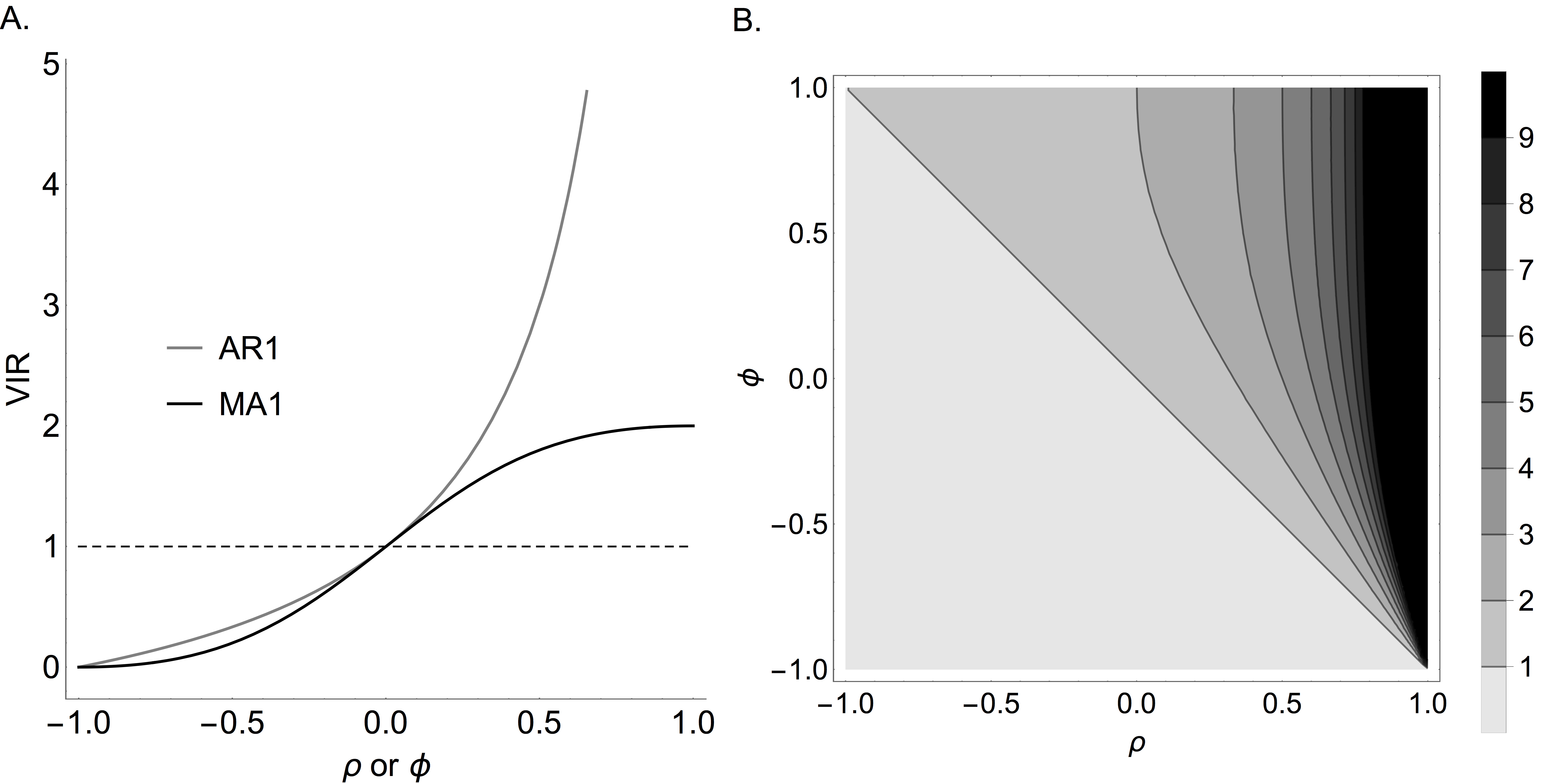}}
	\caption{\textbf{Variance inflation ratios for ARMA processes.} A. shows the VIR for AR(1) and MA(1) processes as a function of their respective parameters; B. shows the VIR for an ARMA(1,1) process.}
	\label{fig:virs}
\end{figure}

The second derivative of eq. (\ref{eq:simple_log_likelihood}) with respect to $\mu$ yields the relevant diagonal element of the FIM,
\begin{equation}\label{eq:fim_mu}
\mathcal{I}_{\mu,\mu} = -\mathbb{E}\left[\frac{\partial^2 \mathcal{L}}{\partial \mu^2}\right] = \frac{T(1-\rho)^2}{\sigma^2}.
\end{equation}
The Cram\'er-Rao lower bound (CRLB) is the asymptotic variance of the maximum likelihood estimator of $\mu$. Because the off-diagonal elements of the FIM are zero, the CRLB is then given by the reciprocal of the RHS of eq. \eqref{eq:fim_mu},
\begin{equation}\label{eq:true_error}
\text{var}(\hat{\mu}) = \frac{\sigma^2}{T(1-\rho)^2}.
\end{equation} 
We next derive the variance of the maximum likelihood estimator of $\mu$ when incorrectly assuming independent errors: ${\epsilon(t)\stackrel{\text{IID}}{\sim} \mathcal{N}(0,\sigma')}$. Under this \textit{false} model, eq. \eqref{eq:true_error} indicates that the variance of maximum likelihood estimators is given by,
\begin{equation}\label{eq:false_error}
\text{var}(\tilde{\mu}) = \frac{\sigma'^2}{T}.
\end{equation}
To meaningfully compare $\text{var}(\tilde{\mu})$ with $\text{var}(\hat{\mu})$, it is necessary to compare estimates of $\sigma'$, the standard deviation of noise for the false error model, with $\sigma$, the standard deviation of $\nu(t)$ in eq. (\ref{eq:ar1}). To do so, we first compute the variance of the (true) AR(1) errors. This can be done by taking the variance of both sides of eq. (\ref{eq:ar1}),
\begin{equation}\label{eq:ar1_var}
\text{var}(\epsilon(t)) = \rho^2 \text{var}(\epsilon(t-1)) + \text{var}(\nu(t)).
\end{equation}
Assuming the error process has a constant variance, eq. (\ref{eq:ar1_var}) can be rearranged to yield:
\begin{equation}
\text{var}(\epsilon(t)) = \frac{\sigma^2}{1-\rho^2}.
\end{equation}
The false error model variance will broadly match the true process variance (otherwise there would be a mismatch between the width of the true and estimated error process) meaning ${\sigma'^2 \approx \sigma^2/(1-\rho^2)}$. Substituting this expression into eq. (\ref{eq:false_error}) and comparing with eq. (\ref{eq:true_error}), we see that true model parameter uncertainty exceeds that obtained from the false model, whenever,
\begin{equation}
\frac{\sigma^2}{T(1-\rho)^2} > \frac{\sigma^2}{T(1-\rho^2)},
\end{equation}
which is true when $0<\rho<1$. The VIR is given by the ratio of the true error uncertainty to that estimated under the false model,
\begin{equation}\label{eq:vif_constant}
\begin{aligned}
\text{VIR}(\rho) &= \frac{1+\rho}{1-\rho}\\
&= 1 + \frac{2\rho}{1-\rho},
\end{aligned}
\end{equation}
which is monotonically-increasing with $\rho$ throughout $0<\rho<1$ (see Figure \ref{fig:virs}A), and ${\lim_{\rho \rightarrow 1} \text{VIR}(\rho) = \infty}$. Intuitively, as autocorrelation increases, each sample conveys less information about the underlying process, and parameter estimates have higher variance. Mischaracterising data as independent, therefore, leads to overly precise estimates.

In our experience, and through the results we present in \S\ref{sec:results}, positive autocorrelation (where $\rho>0$) seems to more commonly occur in systems. If negative autocorrelation does, however, occur, eq. \eqref{eq:vif_constant} indicates that assuming independent noise will produce estimators with inflated variance, and, hence, $\text{VIR}<1$ (see Figure \ref{fig:virs}A).

\subsection{Nonlinear differential equation models}\label{sec:methods_nonliear_ode}
We now consider a model of the form,
\begin{equation}\label{eq:nonlinear_system1}
x(t) = f(t;\theta) + \epsilon(t),
\end{equation}
where, for example, $f(t;\theta)$ is the solution of a nonlinear ODE (or a function of the solution of such an ODE) with univariate parameter $\theta$. As before, the true error process is AR(1) as given by eq. \eqref{eq:ar1}. In \S\ref{sec:methods_ode}, we show that by the same logic as in \S\ref{sec:methods_constant}, the VIR is given by:
\begin{equation}\label{eq:vif_nonlinear}
\text{VIR}(\rho) =  (1-\rho^2)\sum_{t=1}^{T}\left(\frac{\partial f}{\partial \theta}\Bigr|_{t,\theta}\right)^2 / \sum_{t=1}^{T}\left(\frac{\partial f}{\partial \theta}\Bigr|_{t,\theta} - \rho \frac{\partial f}{\partial \theta}\Bigr|_{t-1,\theta}\right)^2.
\end{equation}
If the differential equation solution is linear, its sensitivity is constant, that is, $\partial f/\partial \theta= \text{const}$, and eq. \eqref{eq:vif_constant} for the constant mean model is recovered. If the differential equation has relatively weak nonlinearities, our simulations in \S \ref{sec:results} indicate that eq. \eqref{eq:vif_constant} nonetheless provides a reasonable approximation of eq. \eqref{eq:vif_nonlinear}.

If the model has multiple parameters, so that $\theta$ is a vector, is possible to derive a VIF (see \S\ref{sec:methods_ode_multiple}). However, this expression is not as straightforward to intuit as eq. \eqref{eq:vif_nonlinear}. Indeed, in some of our examples, it is not straightforward to calculate this quantity, and, instead, we approximate the VIF using eq. \eqref{eq:vif_constant}.

Until this point, we have assumed that only the model parameters are unknown, but it is more typical that $\sigma$, $\rho$ and/or the initial state of the system must also be estimated. The results in \S\ref{sec:methods_ode_sigma}\&\ref{sec:methods_ode_rho} show that, since the off-diagonal terms corresponding to $\sigma$ and $\rho$ are zero, that these parameters being unknown does not affect the variances of the $\theta$ estimates. In \S\ref{sec:methods_ode_initial state}, we show that the off-diagonal terms corresponding to $\theta$ and the initial state of the system are generally nonzero: estimates of the model parameters can be correlated with the initial state estimates. This indicates that the exact VIR for model parameters is a less compact expression than eqs. \eqref{eq:vif_constant} or \eqref{eq:vif_nonlinear} when the initial state is unknown. Our results in \S\ref{sec:results}, however, indicate that eq.  \eqref{eq:vif_constant} may nonetheless provide a reasonable approximation in some systems, even for substantially autocorrelated errors.

\subsection{Moving average processes}\label{sec:ma1}
Our results thus far correspond only to AR(1) errors. Other types of autoregressive error processes also exist: one such class is the moving average (MA) processes. In MA processes, the autocorrelation is generally less persistent than for AR processes. The simplest MA process is an MA(1) process, in which a measurement is correlated with its value in the previous period, but not thereafter. This could occur if ephemeral, short-term factors influence consecutive measurements. An MA(1) process can be written:
\begin{equation}\label{eq:ma1}
\epsilon(t) = \nu(t) + \phi \nu(t-1),
\end{equation}
where $\nu(t)\stackrel{\text{IID}}{\sim}\mathcal{N}(0,\sigma)$.

For simplicity of derivation, we revisit the ``constant mean'' model described in \S\ref{sec:methods_constant}, with errors described by an MA(1) process,
\begin{equation}\label{eq:constant_ma1}
x(t) = \mu + \nu(t) + \phi\nu(t-1),
\end{equation}
The steps involved in the calculation of the VIR for the MA(1) case mirror those involved for the AR(1) case and detailed calculations are given in \S\ref{sec:appendix_ma}. The VIR for the $\mu$ parameter of eq. \eqref{eq:constant_ma1} is given by:
\begin{equation}\label{eq:ma1_vir_main}
\text{VIR}(\mu) = 1 + \frac{2\phi}{1+\phi^2},
\end{equation}
meaning the variance of the true model estimator exceeds the false model whenever $\phi>0$ and has a maximum value: $\text{VIR}(\phi=1) = 2$. In \S\ref{sec:appendix_ma}, we describe simulations which we performed to demonstrate the validity of eq. \eqref{eq:ma1_vir_main}. Fig. \ref{fig:mar1_virs} shows the results of these and illustrates that empirical and theoretic VIRs are in good correspondence.

Figure \ref{fig:virs}A demonstrates that, whenever there is positive autocorrelation, $\text{VIR}>1$, meaning that the estimator variance under the true noise model is greater than under the false model. Additionally, if $\rho=\phi>0$ for each of an AR(1) and an MA(1) process, the VIR for the former always exceeds the latter. This makes intuitive sense, since an AR(1) process has greater error persistence meaning that the effects of model misspecification are amplified relative to the more transient MA(1) process.

\subsection{Autoregressive moving-average noise processes}\label{sec:arma}
Noise processes may not neatly fall into either autoregressive or moving average processes; nor need they necessarily be of order 1. In general, noise may be a combination of these two processes, as in the following autoregressive moving-average process formed by combining an AR$(p)$ process with an MA$(q)$ process (termed an ARMA$(p,q)$ process):
\begin{equation}
\epsilon(t) = \rho_1 \epsilon(t-1) + ... + \rho_p \epsilon(t-p) + \nu(t) + \phi_1 \nu(t-1) + ... + \phi_q \nu(t-q). 
\end{equation}
These general processes can be rearranged using the lag operator, ${L a_t = a_{t-1}}$ (see chapter 2 in \cite{harvey1990forecasting} for a discussion of the use and usefulness of lag operators) to:
\begin{equation}\label{eq:polynomial}
\begin{aligned}
\nu(t) &= \frac{1-\rho_1 L - ... - \rho_p L^p}{1 + \phi_1 L + ... + \phi_q L^q} \epsilon(t)\\
&= \frac{\Psi_p(L)}{\Phi_q(L)} \epsilon(t),
\end{aligned}
\end{equation}
where $\Psi_p(L)$ and $\Phi_q(L)$ are shorthands for the corresponding lag operator polynomials. Using eq. \eqref{eq:polynomial}, we can determine the asymptotic variance of the maximum likelihood estimator for $\mu$ in the constant model defined by eq. \eqref{eq:simple_system},
\begin{equation}\label{eq:arma_true}
\text{var}(\hat{\mu}) = \frac{\sigma^2}{T}\frac{\Phi_q(1)^2}{\Psi_p(1)^2}.
\end{equation}
Eq. \eqref{eq:arma_true} gives the variance of the maximum likelihood estimator of $\mu$ when assuming the correct error model. As before, we can also calculate the estimator variance when incorrectly assuming independent Gaussian noise. To do so, requires that we calculate the variance of an ARMA$(p,q)$ process, which for general $p$ and $q$ yields an unwieldy polynomial expansion. Instead, for illustration, we consider the ARMA(1,1) case which has relatively simple variance \cite{harvey1990forecasting} given by:
\begin{equation}
\text{var}(\epsilon(t)) = \frac{1 + \phi^2 + 2\phi \rho}{1-\rho^2}.
\end{equation}
Thus, the VIR is given by,
\begin{equation}\label{eq:arma11_vir}
\text{VIR}(\rho, \phi) = \underbrace{\left(1 + \frac{2\rho}{1-\rho}\right)}_{\text{VIR of AR(1)}}\left(1 + \frac{2\phi (1 - \rho)}{1 + \phi^2 + 2\phi \rho}\right),
\end{equation}
which, as indicated, is the VIR for an AR(1) process multiplied by a factor. This factor exceeds 1 so long as $\phi>0$ and $0<\rho<1$, meaning that the VIR for an ARMA(1,1) process exceeds the VIR for an AR(1) process (and, hence, also that of an MA(1) process) whenever there is positive autocorrelation in terms of both the autoregressive and moving-average terms of the error. This makes intutive sense since, if both constituents of an ARMA(1,1) process cause positive autocorrelation, the combined noise process has even greater autocorrelation.

In \S\ref{sec:appendix_arma11}, we describe simulations we performed to demonstrate the validity of eq. \eqref{eq:arma11_vir}. In Fig. \ref{fig:arma11_virs}, we show the results of these simulations which show that theoretical VIRs are in good correspondence with empirical values.

\section{Applied modelling}
In this section, we first describe in \S\ref{sec:diagnosis} approaches to diagnosing and modelling time series error processes. In \S\ref{sec:model_fitting}, we then describe how to fit these models to data.

\subsection{Determining an appropriate noise process}\label{sec:diagnosis}
When analysing real data, it is generally not straightforward to know what type of measurement process to assume. The canonical assumption is that of IID normal measurements. If a model is fit assuming independent measurements, however, it is possible to test whether the errors -- representing both measurement processes and model discrepancies -- exhibit autocorrelation. Because the errors represent both of these factors, their autocorrelation does not necessarily reflect imperfections in the measurement process. But, if autocorrelation is detected, this forces the analyst to reflect on their chosen measurement model and potentially to refit their model using a more appropriate measurement process. This suggests the following workflow:

\begin{enumerate}
	\item Use an optimiser to fit a model to data. This can be done by targeting either the maximum likelihood parameter values or, alternatively, the Bayesian maximum \textit{a posteriori} (MAP) estimates. We denote the estimated parameter values by $\hat{\theta}$.
	\item Calculate the residuals: $\hat{\epsilon}(t)=x(t) - f(t;\hat{\theta})$. Note these differ from the \textit{true} errors $\epsilon(t)$ since they are obtained using the estimated parameter values rather than the true equivalents.
	\item Calculate the sample autocorrelation function: ${\Gamma(\tau)=\text{cor}(\hat{\epsilon}(t),\hat{\epsilon}(t-\tau))}$ for $\tau\in[1,2,...,\tau_{\text{max}}]$.
	\item If there is evidence of substantial autocorrelation then consider whether this is due to model misspecification or measurement processes. If the former, consider changing the underpinning mechanistic model. If the latter, do a refit assuming an autocorrelated noise model (this fit can either be done via maximisation, for maximum likelihood estimation or MAP estimation; or using, for example, a Markov chain Monte Carlo (MCMC) algorithm for a full Bayesian fitting).
\end{enumerate}

\begin{figure}[ht]
	\centerline{\includegraphics[width=1\textwidth]{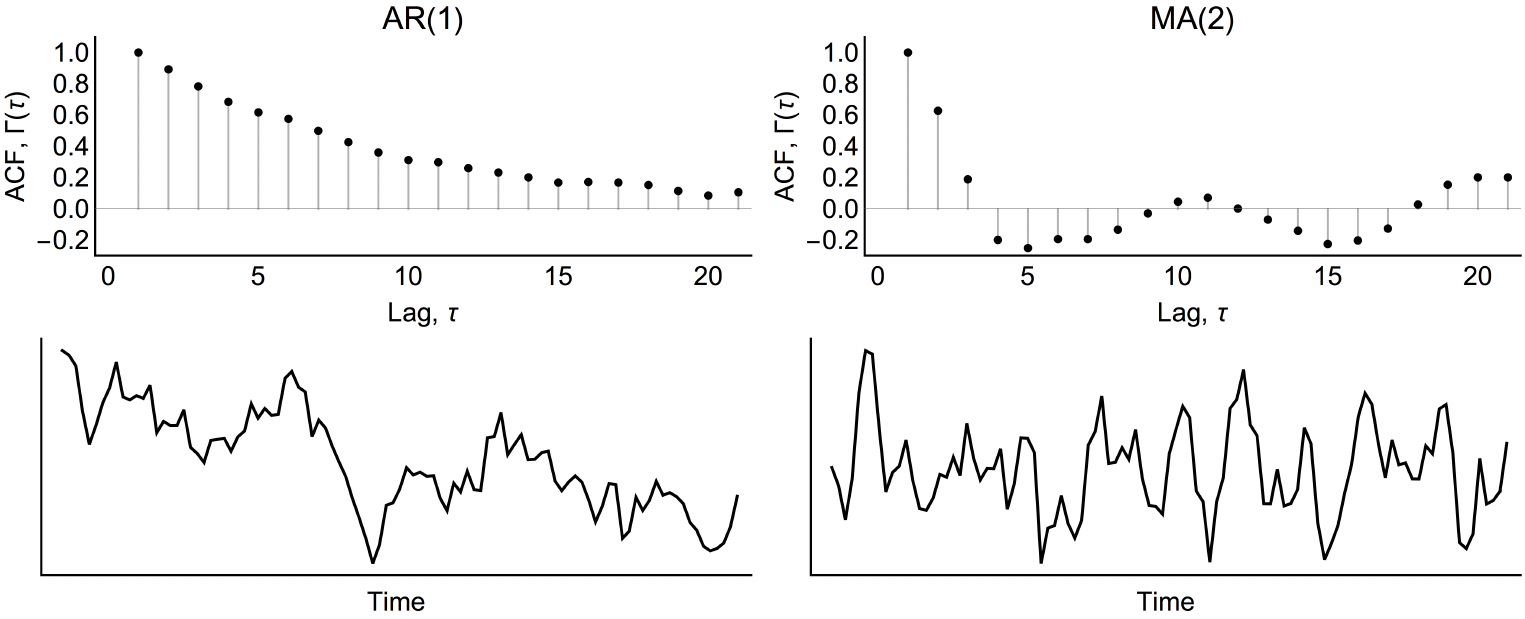}}
	\caption{\textbf{Autocorrelation functions for AR(1) and MA(2) series.} In the bottom panels, we show samples of length 100 of AR(1) and MA(2) processes and, above, their respective sample autocorrelation functions.}
	\label{fig:acfs}
\end{figure}

But if there is evidence of autocorrelated residuals, what autocorrelated noise model should be fit? This depends on the problem at hand but can, as the above suggests, be guided by the sample autocorrelation function of residuals obtained from fitting a model with independent Gaussian errors. For AR(1) processes, the autocorrelation function is \cite{harvey1990forecasting}:
\begin{equation}\label{eq:autocorreltion_ar1}
\Gamma(\tau) = \rho^\tau,
\end{equation} that is, when $|\rho|<1$, an autocorrelation function that decays exponentially with lag (see Figure \ref{fig:acfs}). For MA(1) processes, the autocorrelation function is:
\begin{equation}\label{eq:autocorrelation_ma1}
\Gamma(\tau)= 
\begin{cases}
\phi,& \text{if } \tau= 1\\
0,              & \text{otherwise}.
\end{cases}
\end{equation}
So, for MA(1) processes, substantial autocorrelation occurs only at the first lag. More generally, for MA($q$) processes, autocorrelation exists until the $q$th lag (see Figure \ref{fig:acfs}). Indeed, whenever $|\rho|<1$, it is possible to use the Koyck transformation to rewrite an AR(1) process as an MA($\infty$) process (with MA coefficients exactly mirroring the autocorrelations given in eq. \eqref{eq:autocorreltion_ar1}), which provides some intuition for the interrelation between these two types of process \cite{harvey1990forecasting}.

\subsection{ARMA and other time series models}\label{sec:methods_STS}
Choosing an ARMA error process that mirrors the autocorrelation patterns seen in the residuals provides a somewhat automated way of deciding on a noise model and, essentially, follows the approach forged by Box and Jenkins in their pathbreaking 1970s book (recent edition: \cite{box2015time}). This framework is, by no means, the only workflow followed, since applied time series modelling is, actually, a much broader church.  An alternative popular approach falls under the banner of ``structural time series (STS)'' or ``state-space'' modelling, championed originally by Harvey for econometric time series \cite{harvey1990forecasting}. In this philosophy, a time series is built up from various latent (i.e. not directly observed) components that represent characteristics of the series. For example, a series may be decomposed into stochastic time trends and cyclical components.

The STS approach is more model-driven and aims to decompose a series into understandable components. The STS framework is also naturally able to handle series that are non-stationary, where the probability distributions governing quantities like the mean and variance of the process vary over time. In the Box-Jenkins approach, by contrast, any nonstationarity is treated first by differencing the series, that is, via the operator, $\Delta_s y_t = y_t - y_s$, then by fitting an ARMA model to the transformed series -- this combined process of differencing followed by fitting ARMA models is termed autoregressive integrated moving-average process (ARIMA) modelling.

Since both types of time series analysis -- ARMA and STS -- are used in practice, we do not suggest a single path here. In the two real data examples in \S\ref{sec:results}, we initially follow Box-Jenkins and examine how well different ARMA models fit the residual series using the Akaike Information Criterion. This provides us with a guide as to whether models allowing autocorrelation better fit the data and hints as to which alternative models should be fitted -- particularly as, in our examples, it is feasible that measurement apparatus imperfections could lead to residual autocorrelation.

\subsection{Model fitting}\label{sec:model_fitting}
When an appropriate error process has been chosen using the framework described in \S\ref{sec:diagnosis}, it is necessary to fit the model to data. For ARMA processes, there are essentially two ways to fit such models to data: the first uses the generative process model to write down a conditional likelihood; the second, and more general approach, uses Kalman filters, which provide an efficient means to calculate likelihoods. An additional benefit of Kalman filters is that they can also handle STS-type models (see \S\ref{sec:methods_STS}). Here, we describe how the first, and simpler, of these approaches can be used to fit an ODE model with ARMA(1,1) errors. The equivalent Kalman filter approach is provided in \S\ref{sec:kalman}. In both cases, we suppose that the measurement equation for a univariate system observable is determined by the following system:
\begin{equation}\label{eq:arma11}
\begin{aligned}
x(t) &= f(t;\theta) + \epsilon(t)\\
\epsilon(t) &= \rho \epsilon(t-1) + \nu(t) + \phi \nu(t-1),
\end{aligned}
\end{equation}
where, as in previous cases, $\nu(t)\stackrel{\text{IID}}{\sim}\mathcal{N}(0,\sigma)$.

To determine the likelihood for this model, we assume that the first two terms $\nu(1)=0$ and $\nu(2)=0$: this is known as a ``conditional likelihood'' approach because we condition on initial values of processes\footnote{This approach follows the discussion in chapter 5.6 of \cite{hamilton1994time}.}. (Alternatively, rather than directly specifying $\nu(1)$ and $\nu(2)$, in a Bayesian framework, these can be set priors, allowing them to potentially take non-zero values.) For a given value of $\theta$, the error can be directly calculated using $\epsilon(t) = x(t) - f(t;\theta)$. Putting these together, we obtain:

\begin{equation}
\nu(t) = \epsilon(t) - \rho \epsilon(t-1) - \phi \nu(t-1), \; \forall t>2.
\end{equation}

Thus, the log-likelihood for this model is given by,

\begin{equation}\label{eq:conditional_loglikelihood}
\mathcal{L} = -\frac{T-2}{2} \text{log}\; 2\pi - \frac{T-2}{2} \text{log}\;\sigma^2 - \frac{1}{2\sigma^2} \sum_{t=3}^{T} \nu(t)^2.
\end{equation}

The results shown in \S\ref{sec:results} of this paper were generated assuming such a conditional likelihood approach.

\section{Results}\label{sec:results}
Here, we present results that illustrate the importance of assessing the validity of independent measurements and the consequences of failing to account for these measurement imperfections, when present. In \S\ref{sec:logistic}, we first use synthetic data generated from a logistic model. In \S\ref{sec:herg}, we then use real data from cardiac electrophysiology experiments. In \S\ref{sec:electrochemistry}, we model outputs from electrochemistry experiments.

\subsection{Logistic model}\label{sec:logistic}
In this section, we use a simple ODE model to demonstrate how failing to account for autocorrelated measurements can lead to overly confident estimates; it also shows how mistakenly assuming independent measurements leads to more variable estimates. Here, we use the logistic model, which is a univariate ODE, with solution determined from,
\begin{equation}\label{eq:logistic}
\frac{dx(t)}{dt} = r x(t) \left(1 - \frac{x(t)}{\kappa}\right),
\end{equation}
where $r>0$ is a parameter determining the initial exponential growth rate, and ${\kappa=\lim_{t \to \infty} x(t)}$ is the carrying capacity; $x(0) >0$ is the initial output value. The logistic model is common in mathematical biology, where it is typically used to describe resource-limited growth: imagine bacteria dividing on an agar plate -- initially, bacteria have access to much resource, and the population density grows fast; later, once food becomes scarce, growth slows and the population eventually reaches a maximum size.

In our experiments, we generated $x(t)$ using $r=0.5$, $\kappa=50$, and $x(0)=1$. We then generated observations $y(t)=x(t) + \epsilon(t)$ and used AR(1) errors, $\epsilon(t)$, as described by eq. \eqref{eq:ar1}, where we fixed $\sigma=1$ and used five $\rho$ values between 0.8-0.975 to generate synthetic datasets. For each $\rho$ value, we generated a dataset consisting of 2000 equally spaced observations between $t=0$ and $t=20$. Ten such replicate datasets were generated for each $\rho$ value. For each of these replicates, we fit two statistical models: the correct one, which assumes AR(1) errors; the other, with IID Gaussian errors. For both models, we estimated $r, \kappa, x(0)$ and $\sigma$; for the AR(1) model, we also estimated $\rho$. For the AR(1) model, we calculated the likelihood using the \textit{generative model log-likelihood} approach described in \S\ref{sec:model_fitting}. The priors that we use for each parameter are shown in Table \ref{table:logistic_priors}. The ODE was solved using Stan's Runga-Kutta 4-5 solver \cite{carpenter2016stan}. These models are fit using Stan's NUTS MCMC algorithm \cite{hoffman2014no} with 2000 iterations across each of 4 chains, with 1000 initial iterations discarded as warm-up. In all cases, $\hat{R}<1.01$ for all model parameters diagnosing MCMC convergence \cite{gelman1992inference}.

\begin{figure}[ht]
	\centerline{\includegraphics[width=1\textwidth]{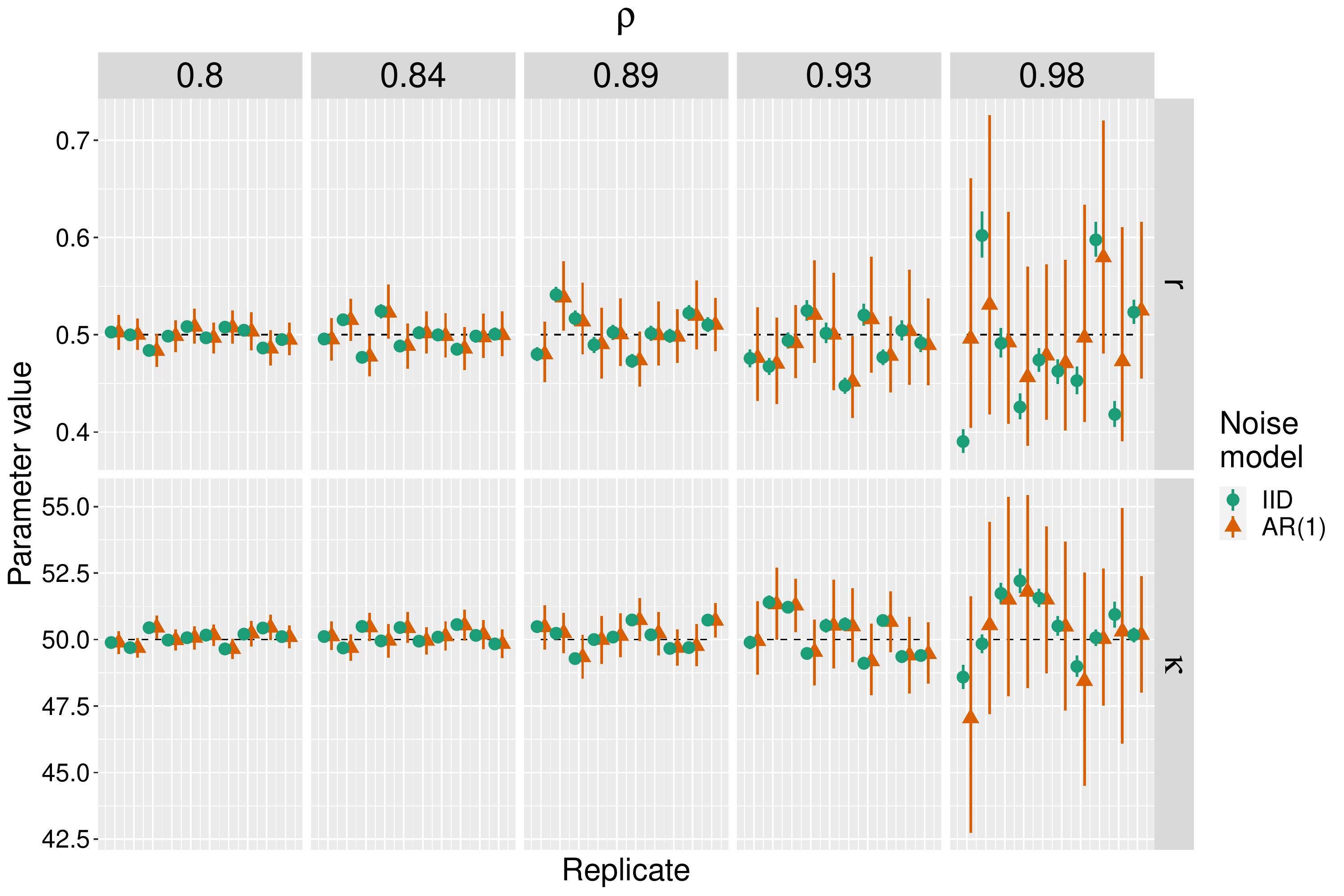}}
	\caption{\textbf{Logistic model: posteriors.} Columns show results for each value of $\rho=\text{cor}(\epsilon(t), \epsilon(t-1))$ used to generate synthetic datasets as described in \S\ref{sec:logistic}. Rows show posteriors for each of $r$ and $\kappa$. Within each panel, we show posteriors from both the IID and AR(1) noise models; in each (of 10) replicates, both noise models were fitted to the same synthetic data which are shown as pairs of IID and AR(1) posteriors. Upper and lower whiskers represent 2.5\% and 97.5\% posterior quantiles; points represent posterior medians. Dashed lines show true parameter values.}
	\label{fig:logistic_posteriors}
\end{figure}

In Fig. \ref{fig:logistic_posteriors}, we show summaries of the posterior distributions for the logistic model parameters for both the IID and AR(1) models fitted to each of the replicate datasets. The columns show results for different values of $\rho$; the rows show separate results for $r$ and $\kappa$ in eq. \eqref{eq:logistic}. Within each panel, we show the IID and AR(1) posteriors for each replicate dataset.

\begin{figure}[ht]
	\centerline{\includegraphics[width=1\textwidth]{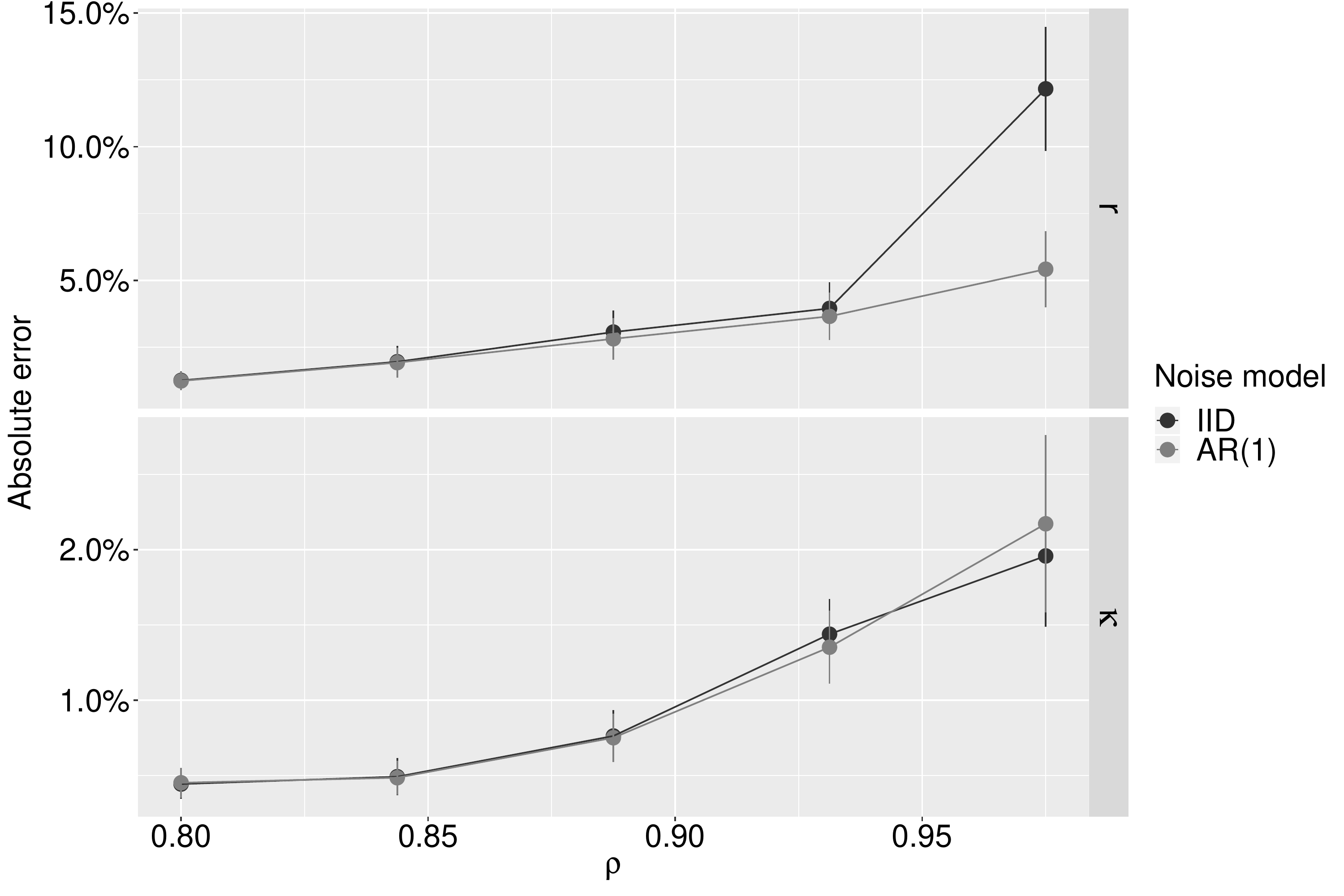}}
	\caption{\textbf{Logistic model: point estimate errors.} The horizontal axis indicates the values of $\rho$ used to generate synthetic datasets using AR(1) errors as described in \S\ref{sec:logistic}. The vertical axis shows the absolute percentage error in estimating the true parameter values from $r=0.5$ (top panel) and $\kappa=50$ (bottom panel) across both the IID and AR(1) noise models. The upper and lower whiskers of the ranges represent 2.5\% and 97.5\% quantiles in absolute percentage error; the points represent medians.}
	\label{fig:logistic_errors}
	\end{figure}

We focus first on point estimates of the parameter values (the points and triangles in Fig. \ref{fig:logistic_posteriors}). Across the two model parameters and both noise models, the point estimates become more variable as $\rho$ increases. Yet, over each set of replicates, the estimates appear relatively unbiased, with point estimates as likely to overestimate the true values as to understate them. The extent of variation, however, differs between the two models, and for 71\% of replicates, the point estimate from the AR(1) model was closer to the true parameter value than the equivalent from the IID model. In Fig. \ref{fig:logistic_errors}, we quantify this by calculating the absolute percentage error in estimating each parameter value across all replicates at a given value of $\rho$ for both noise models. This shows that the predictive errors for the logistic growth parameter, $r$, were between 1\%-12\% over all $\rho$ values considered; the errors for the carrying capacity, $\kappa$, were, in general, lower (at around 0.5\%-2\%). This difference in accuracy is likely due to the somewhat narrower range of times when the model solution is sensitive to small changes in $r$ as opposed to $\kappa$.  Fig. \ref{fig:logistic_errors} also shows that as $\rho$ increases, both models get worse at estimating the true parameter; for $r$, the AR(1) model, however, does better on average than the IID one; for $\kappa$, both models perform similarly in terms of average error.

\begin{figure}[ht]
	\centerline{\includegraphics[width=0.8\textwidth]{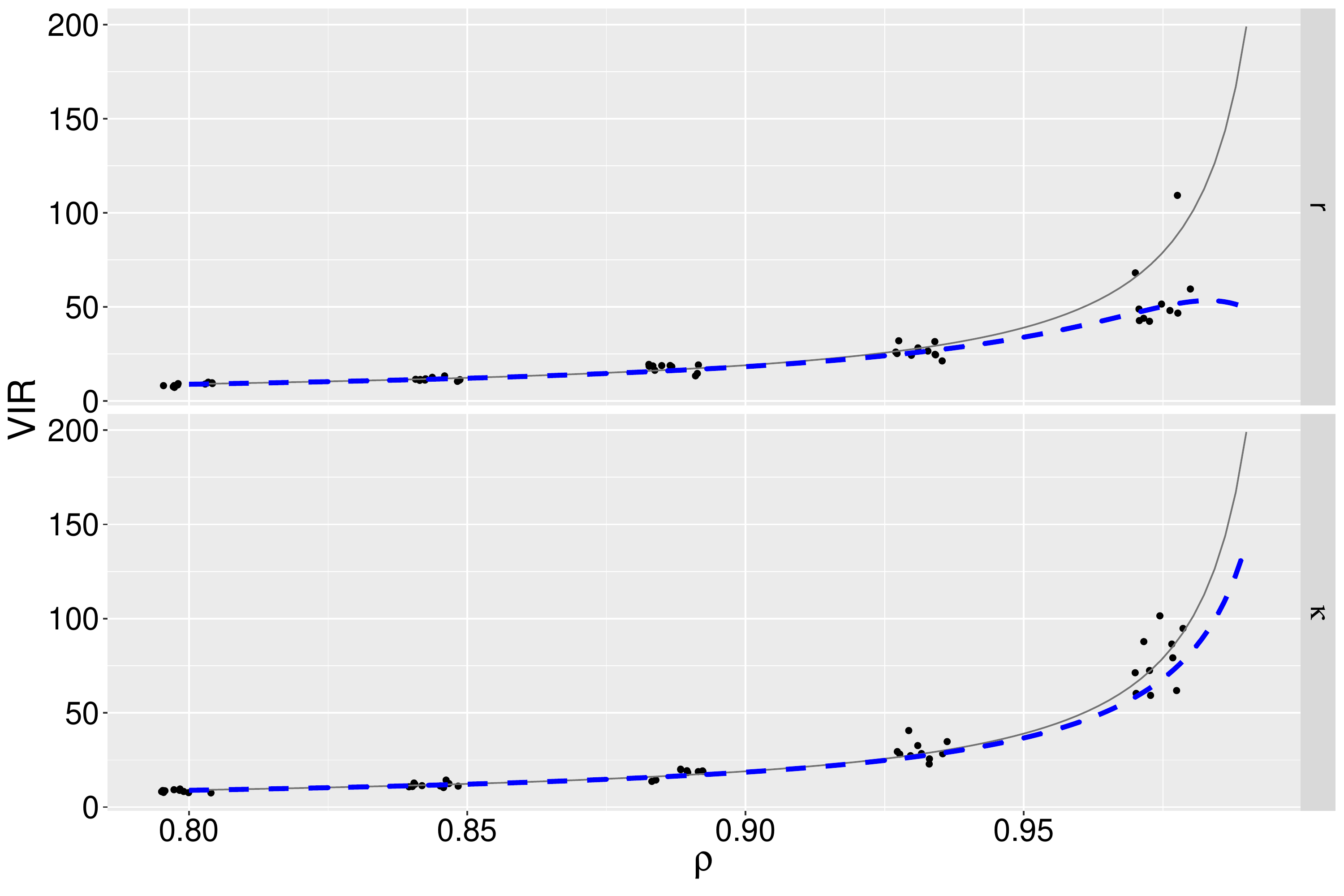}}
	\caption{\textbf{Logistic model: variance inflation ratios.} The rows show results for the two logistic model parameters, $r$ and $\kappa$. The vertical axis shows the estimated variance inflation ratios (VIRs) at each $\rho$ value, which is the ratio of the AR(1) posterior variance to that of the IID model for each replicate (points). The lines show theoretically predicted VIRS: the blue-dashed line shows a more accurate VIR, accounting for uncertainty in the initial state (see \S\ref{sec:methods_ode_initial state}), and the grey line plots eq. \eqref{eq:vif_constant} which ignores this uncertainty and treats the ODE solution as a linear model. Note, that horizontal jitter has been added to the points.}
	\label{fig:logistic_virs}
\end{figure}

We next examine the uncertainty in estimates (the whiskers in Fig. \ref{fig:logistic_posteriors}). Across the two model parameters and both noise models, the posterior uncertainties widen as $\rho$ increases. The extent to which they increase in width differs across both noise models, however, with the AR(1) uncertainties widening more acutely with changes in $\rho$. Indeed, for each replicate, we can calculate the ratio of the posterior variance for the AR(1) model to the IID model -- in effect, estimating a VIR in each case -- which we show in Fig. \ref{fig:logistic_virs}. The two rows here both show how the VIRs for each logistic model parameter increase along with $\rho$. To illustrate how our theory predicts this change, we also plot the theoretical VIR (blue dashed lines; see \S\ref{sec:methods_ode_initial state}) and the more approximate VIR which assumes the function is constant (eq. \eqref{eq:vif_constant}; grey lines). Note that both VIRs plotted are somewhat approximate since they are derived from considering maximum likelihood estimates for an unbounded parameter, which is an approximation in this case since both $r$ and $\kappa$ are bounded below at zero, and we perform Bayesian inference using Gaussian priors. The theoretical results nonetheless capture well how the VIRs change with $\rho$, and eq. \eqref{eq:vif_constant} performs similarly to the more accurate result until the degree of autocorrelation is very high.

\begin{figure}[ht]
	\centerline{\includegraphics[width=1\textwidth]{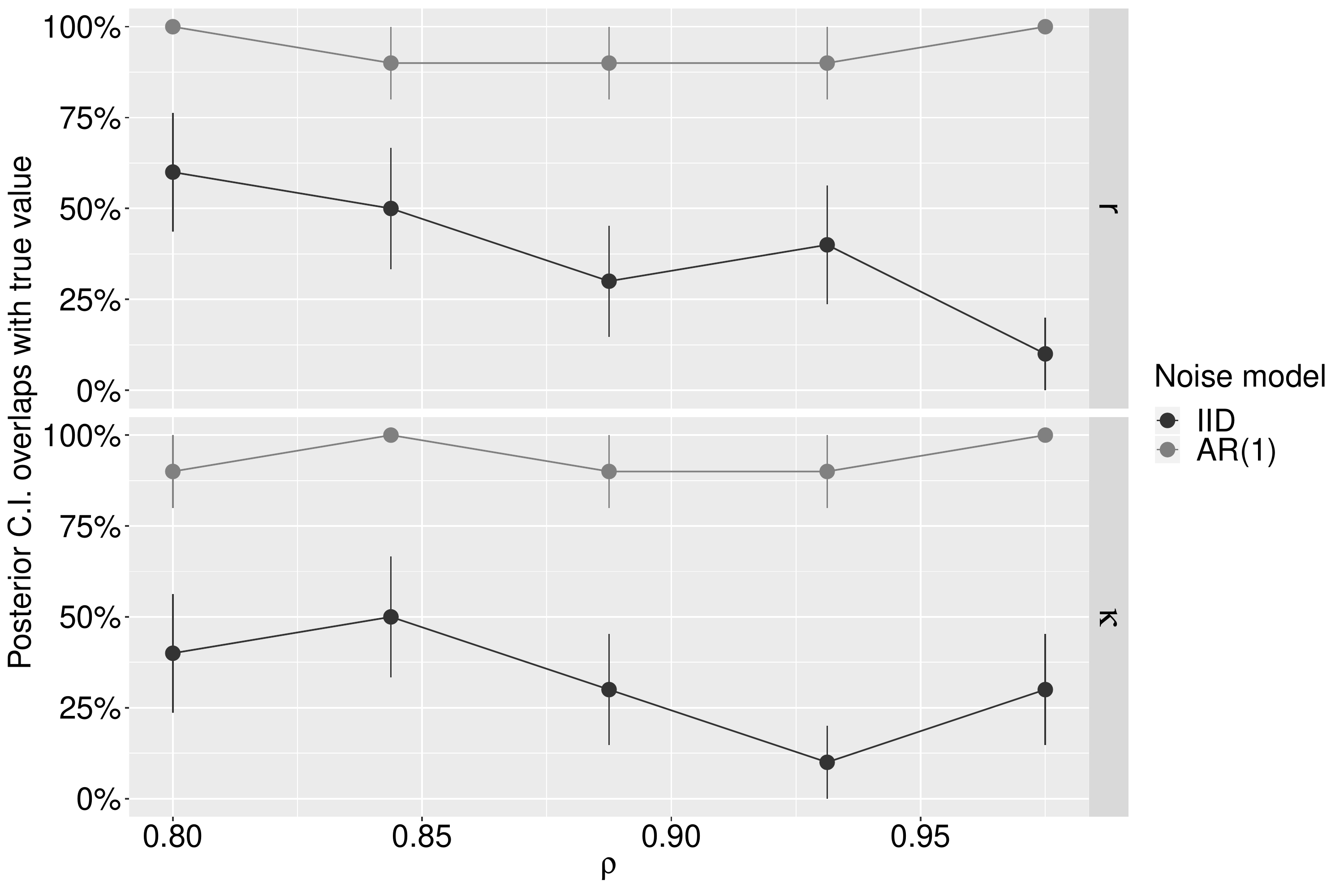}}
	\caption{\textbf{Logistic model: estimate overlap with true parameter values.} The horizontal axis indicates the values of $\rho$ used to generate synthetic datasets using AR(1) errors as described in \S\ref{sec:logistic}. The vertical axis shows the percentage of replicates where the 95\% posterior interval includes the true parameter value for $r=0.5$ (top panel) and $\kappa=50$ (bottom panel) across both models. Points indicate the mean percent of replicates where the posterior interval includes the true parameter values; the upper and lower whiskers show the standard error in the mean.}
	\label{fig:logistic_overlap}
\end{figure}

Finally, we examine how frequently the 95\% posterior interval for the IID and AR(1) model posteriors encompass the true parameter value: we call these cases ``successes''. In Fig. \ref{fig:logistic_overlap}, we show the percentage of successes for $r$ and $\kappa$ at each value of $\rho$ examined. Overall, this shows that the AR(1) posterior intervals more frequently encompass the true parameter value than the IID model. Indeed, across all values of $\rho$ investigated, the maximum success percentage for $r$ was 60\% for the IID model and 100\% for the AR(1) model (the results were qualitatively similar, albeit of different magnitudes for $\kappa$). Additionally, as $\rho$ increased, the frequency of success decreased for both parameters in the IID model; in all cases, the AR(1) model success frequencies did not change directionally with $\rho$.

 Overall, our results show that using an inappropriate noise model results in more variable point estimates of parameters and uncertainties that are less reliable. This result has long been appreciated in time series regression analyses, where methods like Generalised Least Squares -- which essentially attempt to account for the structure of the noise -- are commonly used when errors appear to deviate from IID Gaussian \cite{wooldridge2015introductory}.

\subsection{Cardiac electrophysiology model}\label{sec:herg}
We next examine a real dataset collected from experiments in cardiac electrophysiology on the \textit{human Ether-\`{a}-go-go-Related Gene} (hERG) ion channel. These datasets are published with these journal articles: \cite{lei_rapid_2019-1,lei_rapid_2019-2}. In the experiments, current from the hERG channel, which is often referred to as the rapid delayed rectifier potassium current, $I_\text{Kr}(t)$, is measured under a time-varying voltage stimulus $V(t)$. The same laboratory experiment was conducted on five different cells, and we fit to each of these datasets separately, producing five sets of estimates.

Here, we model the current response of the hERG channel to this stimulus using an ODE model in the flavour of Hodgkin \& Huxley's (HH) landmark study \cite{hodgkin1952quantitative}. This model contains two HH-style gating variables (`activation' $a$ and `recovery' from inactivation $r$) and a standard Ohmic expression,
\begin{equation}\label{eq:ikr}
I_\text{Kr}(t) = g_\text{Kr} \cdot a(t) \cdot r(t) \cdot (V(t) - E_\text{K}),
\end{equation}
where $g_\text{Kr}$ is the maximal conductance, and $E_\text{K}$ is the reversal potential (Nernst potential) for potassium ions, which can be calculated directly from potassium concentrations using the Nernst equation. The voltage stimulus is a complicated ``staircase-like'' function with no simple closed form: see \cite{lei_rapid_2019-1} for further description. The gates $a$ and $r$ are governed by the ODEs:
\begin{align*}
\frac{da}{dt} &= \frac{a_{\infty} - a}{\tau_{a}}, &
\frac{dr}{dt} &= \frac{r_{\infty} - r}{\tau_{r}}, \\
a_{\infty} &= \frac{k_1}{k_1 + k_2}, &
r_{\infty} &= \frac{k_4}{k_3 + k_4}, \\
\tau_{a} &= \frac{1}{k_1 + k_2}, &
\tau_{r} &= \frac{1}{k_3 + k_4},
\intertext{where}
k_1 &= p_1 \exp(p_2 V),    & k_3 &= p_5 \exp(p_6 V), \\
k_2 &= p_3 \exp(-p_4 V),   & k_4 &= p_7 \exp(-p_8 V).
\end{align*}

The model has 9 positive parameters to be inferred from the experimental data: maximal conductance $g_\mathrm{Kr}$ and kinetic parameters $p_1, p_2, p_3, \cdots, p_8$. The initial conditions of the system were assumed to be: $a(0) = 0$ and $r(0) = 1$ and the system was solved for 100 seconds at $V=-80mV$ before running the staircase protocol.

Here, we assume that the measured current differs from the true current and is described by $I(t) = I_\text{Kr} + \epsilon(t)$, where $\epsilon(t)$ is an error process that can either be IID Gaussian, $\epsilon(t)\stackrel{\text{IID}}{\sim} \mathcal{N}(0,\sigma)$, or described by an autoregressive process.

First, we use optimisation to determine whether there is evidence of autocorrelation in the errors. To do so, we maximise the posterior assuming IID noise and from this to obtain a residual series. For optimisation, we used CMA-ES \cite{hansen2016cma}, a derivative-free optimiser, as implemented in PINTS \cite{Clerx2019Pints} following previous work \cite{lei_rapid_2019-1,lei_rapid_2019-2}. In Fig. \ref{fig:herg_acfs}, we plot the sample autocorrelation function for the residuals for each of the cells, which illustrates strong and persistent autocorrelation, characteristic of autoregressive processes. Across all cells, the estimated 1st order residual autocorrelation was between 0.57 and 0.83.

We then compared the fit of the residual series to a range of ARMA processes: MA(1), AR(1) and ARMA(1,1), all of which could reasonably represent experimental artefacts: for example,  series resistance and leakage currents \cite{lei_accounting_2019}. For each cell, we calculated the Akaike Information Criterion (AIC) for a range of ARMA($p$,$q$) processes (where a lower AIC indicates a better fitting model \cite{akaike1974new}). The best ARMA model varied by cell and optimal $p$ was between 1-5 and $q$ from 2-5 (see Fig. \ref{fig:herg_arma}). In Fig. \ref{fig:herg_aics}, we show the result of these comparisons. Each panel of this figure corresponds to a cell. In each panel, we show the percentage difference between the AICs of each other process to the best fitting ARMA model (``Min AIC''). In all cases, this shows that the IID Gaussian model is bettered by models encompassing autocorrelation. It also shows that the models incorporating autoregressive terms outperformed the MA(1) model. In all cases, the ARMA(1,1) model produced a similar quality fit to the best model. Because of this, we decided only to attempt to perform Bayesian inference for the full model using the more parsimonious ARMA(1,1) noise compared to the best fitting ARMA($p$,$q$) process.

To perform Bayesian inference, we used MCMC sampling for the IID, AR(1) and ARMA(1,1) noise models. For the sampling, we used population MCMC, which runs a series of chains at different ``temperatures'' \cite{jasra2007population}, using the default PINTS \cite{Clerx2019Pints} algorithm settings. For each noise model and each of five cells, we ran four Markov chains with 150,000 iterations on each, with the first 50,000 of these discarded as warm-up; the draws were thinned by a factor of 10 after sampling.

Whilst the ARMA(1,1) model was the best fit to the residuals, we could not achieve convergence with this model despite trying a range of informative priors on noise parameters. The difficulty of performing Bayesian inference for ARMA models has been noted before \cite{kleibergen1997bayesian}. Because of this, we present results only for the IID and AR(1) models, which had $\hat{R} < 1.1$ for all parameters. The priors specified for these two models are shown in Table \ref{table:herg_priors}.

\begin{figure}[ht]
	\centerline{\includegraphics[width=1\textwidth]{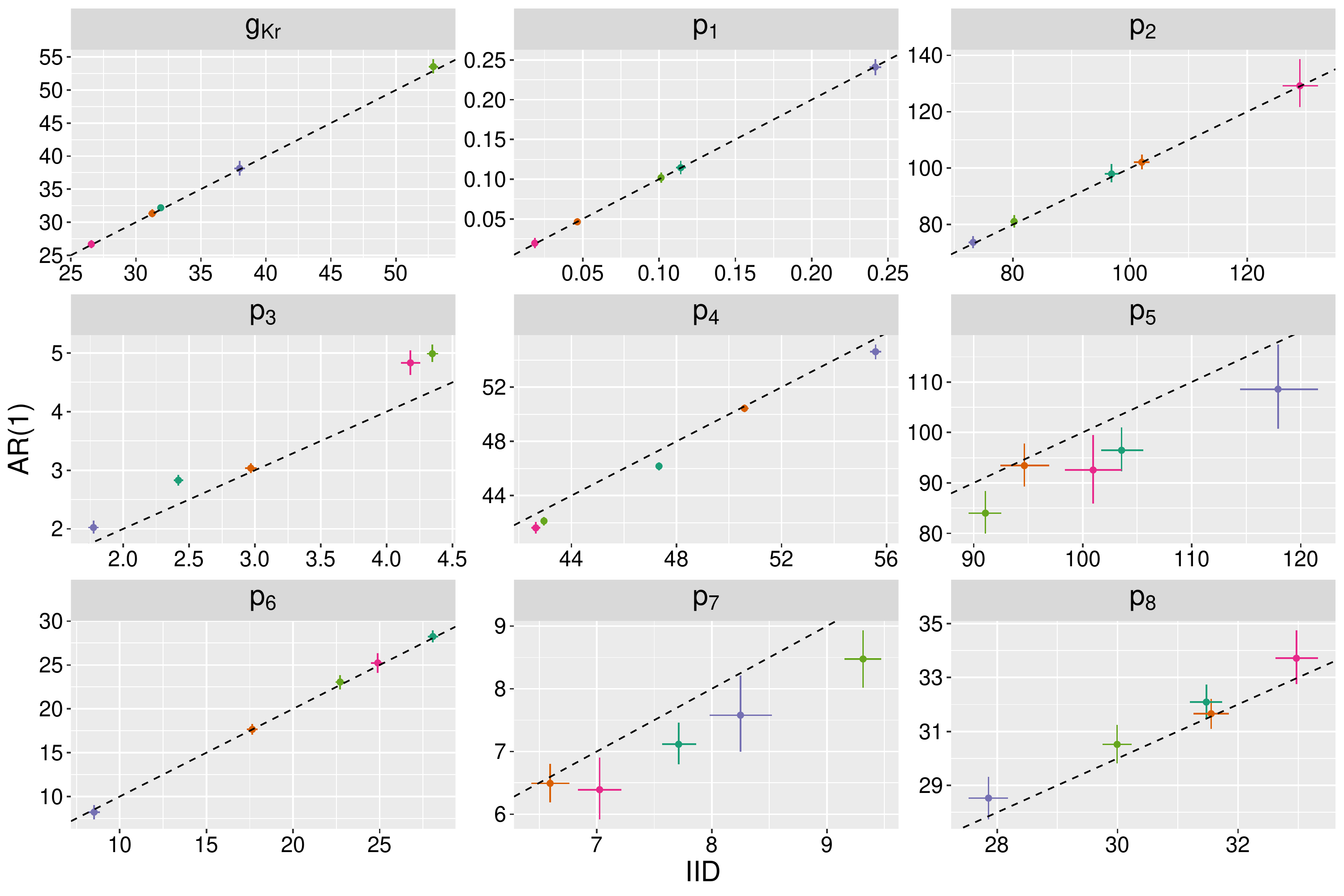}}
	\caption{\textbf{hERG model: posteriors.} The horizontal axis shows the parameter values estimated from the IID noise model; the vertical axis shows the same for the AR(1) noise model. Points show posterior medians; whiskers indicate the 97.5\% and 2.5\% posterior quantiles. Colours indicate the experimental replicate (i.e. the cell on which experiments were performed). The dashed line shows the $y=x$ line. Note, that the values for $g_{Kr}$ have been scaled down by a factor of 1000; the $p_3$ values have been scaled up by a factor of 100.}
	\label{fig:herg_posteriors}
\end{figure}

\begin{figure}[ht]
	\centerline{\includegraphics[width=1\textwidth]{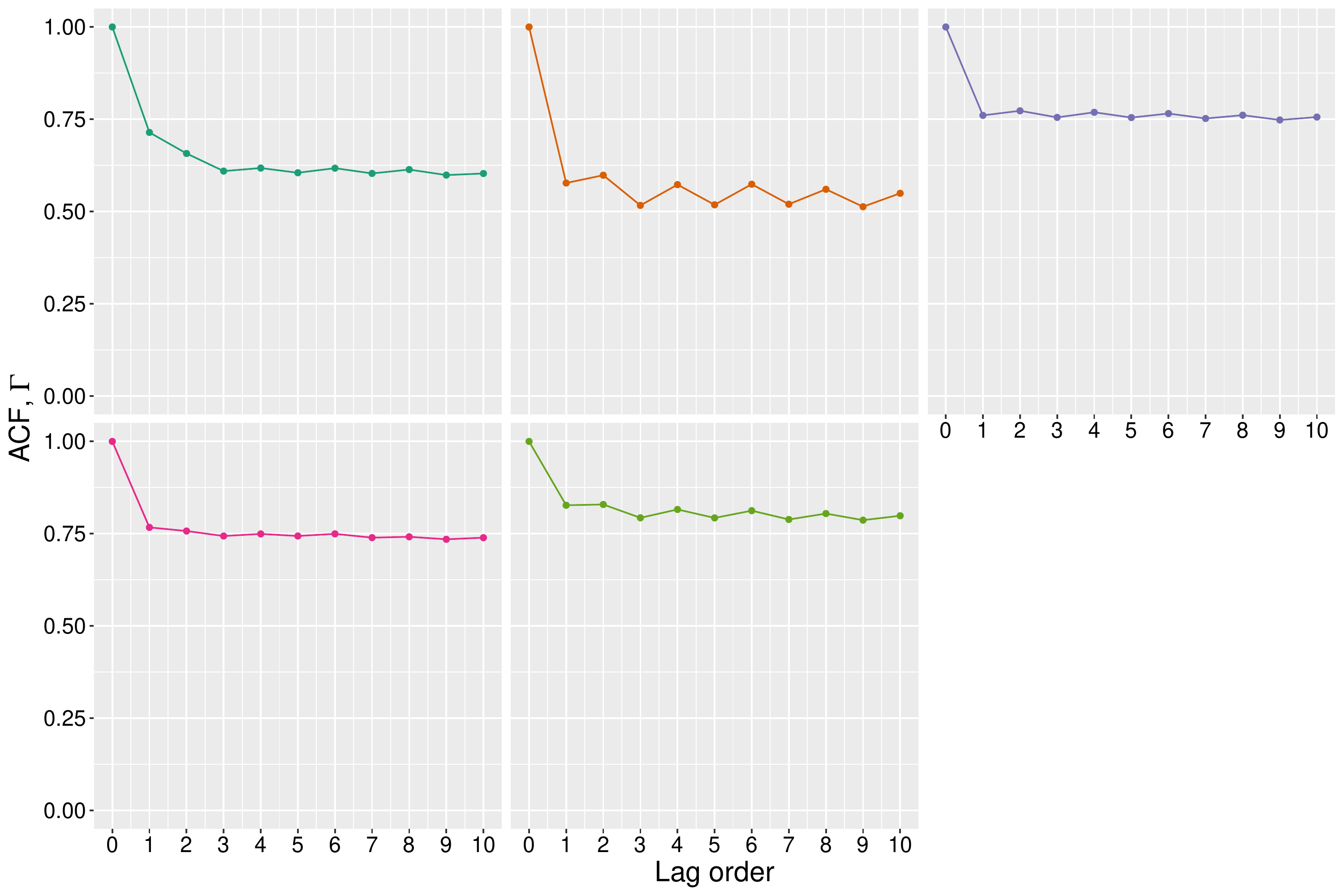}}
	\caption{\textbf{hERG model: sample autocorrelation functions.} Each panel shows estimates for a given experimental replicate: colours indicate the experimental replicate (i.e. the cell on which experiments were performed) and correspond with those shown in Fig. \ref{fig:herg_posteriors}.}
	\label{fig:herg_acfs}
\end{figure}

\begin{figure}[ht]
	\centerline{\includegraphics[width=1\textwidth]{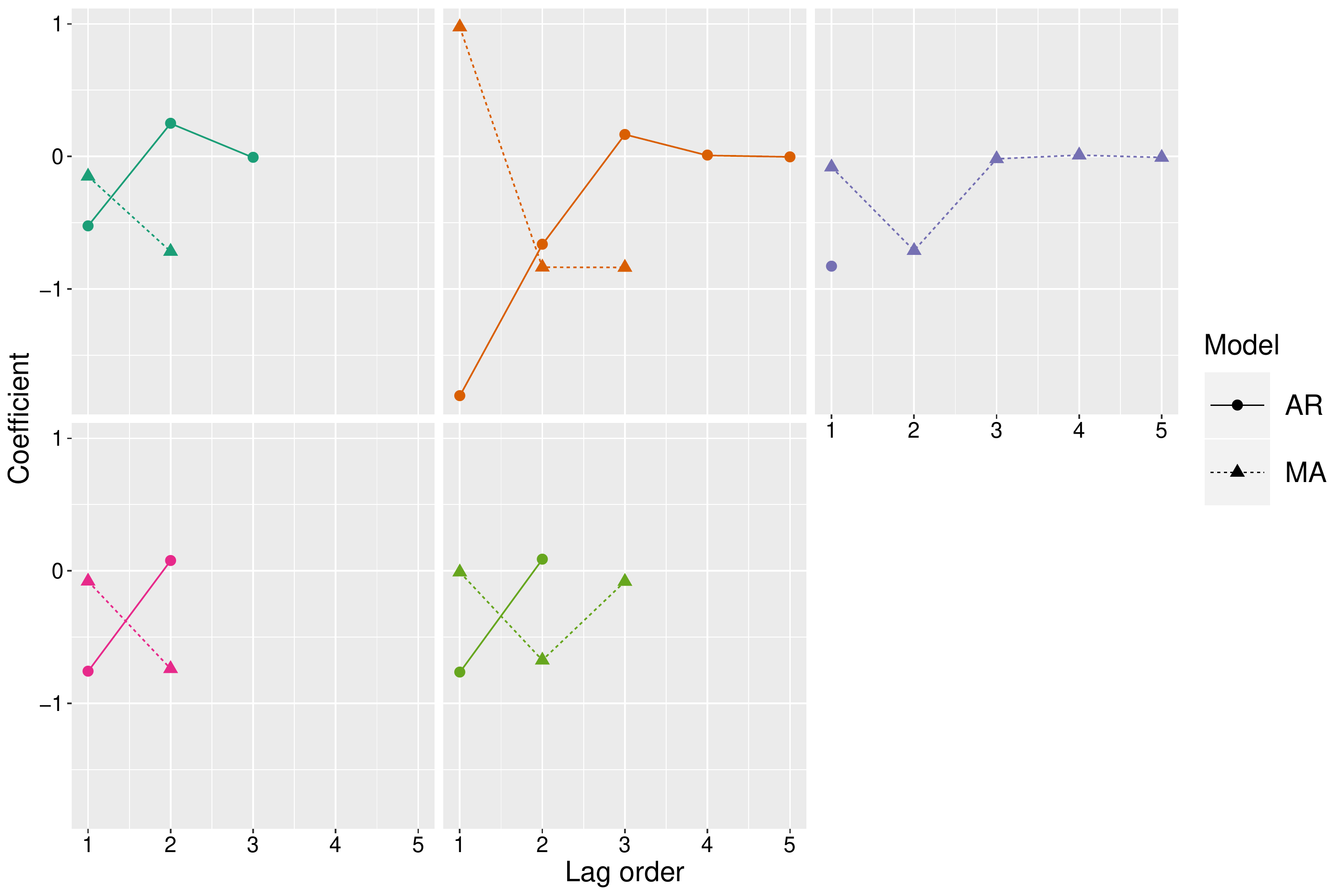}}
	\caption{\textbf{hERG model: optimal ARMA coefficients.} Each panel shows estimates for a given experimental replicate: colours indicate the experimental replicate (i.e. the cell on which experiments were performed) and correspond with those shown in Fig. \ref{fig:herg_posteriors}. The lines show the estimated AR (solid line with points) and MA (dashed line with triangles) coefficients as determined by AIC when fitting ARMA($p$,$q$) processes to the residual series resultant from a single optimisation which assumed IID noise. Note, that if values are missing from a given lag order (e.g. for lag order 3 upwards for the pink cell), this indicates that the optimal model was of lower order.}
	\label{fig:herg_arma}
\end{figure}

\begin{figure}[ht]
	\centerline{\includegraphics[width=1\textwidth]{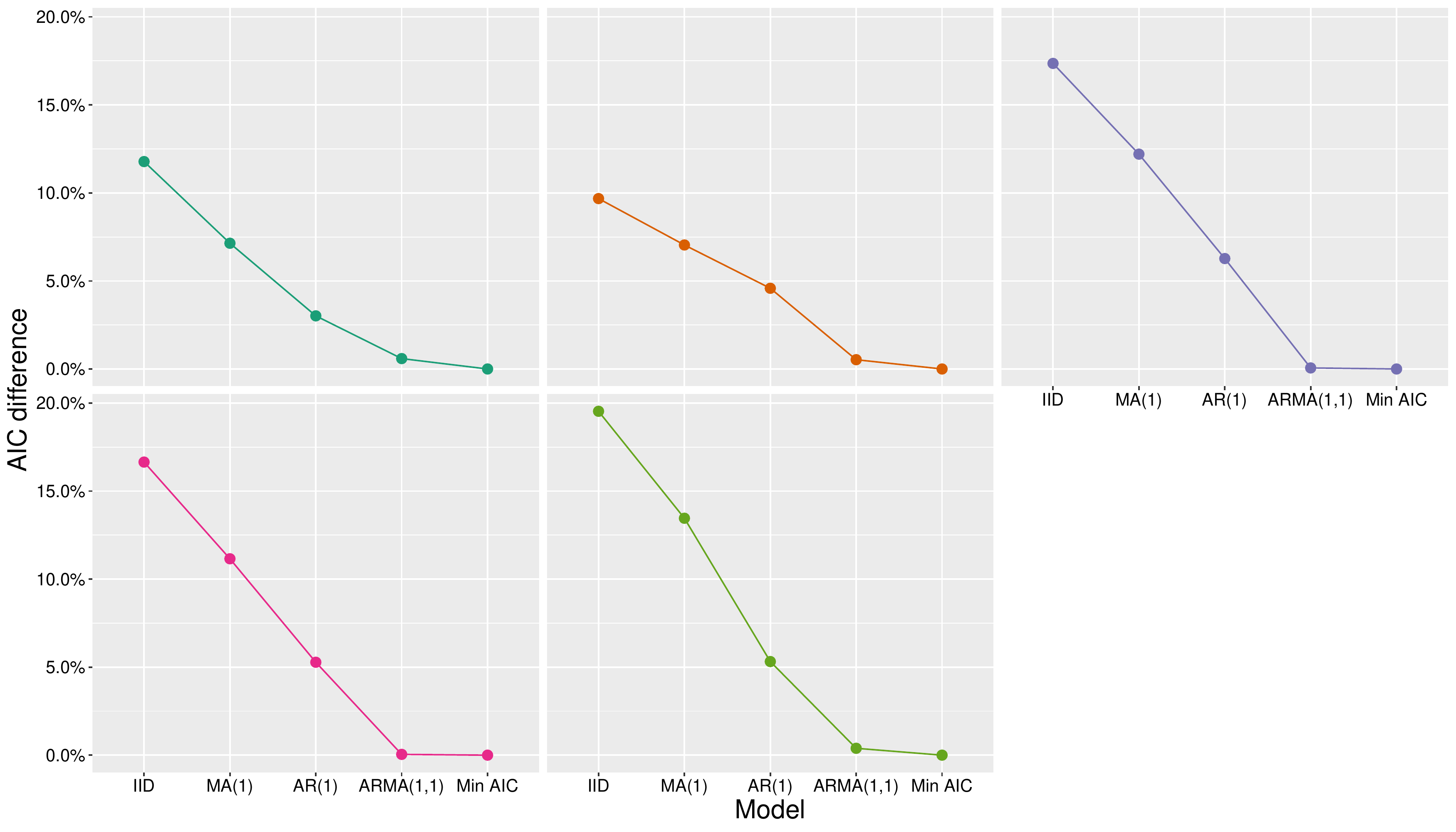}}
	\caption{\textbf{hERG model: AICs.} Each panel shows estimates for a given experimental replicate: colours indicate the experimental replicate (i.e. the cell on which experiments were performed) and correspond with those shown in Fig. \ref{fig:herg_posteriors}. The vertical axis shows the percentage difference between the AIC of each model (indicated on horizontal axis) compared to the best fitting model (``Min AIC'') when fitted to the residual series resultant from a single optimisation which assumed IID noise.}
	\label{fig:herg_aics}
\end{figure}

\begin{figure}[ht]
	\centerline{\includegraphics[width=1\textwidth]{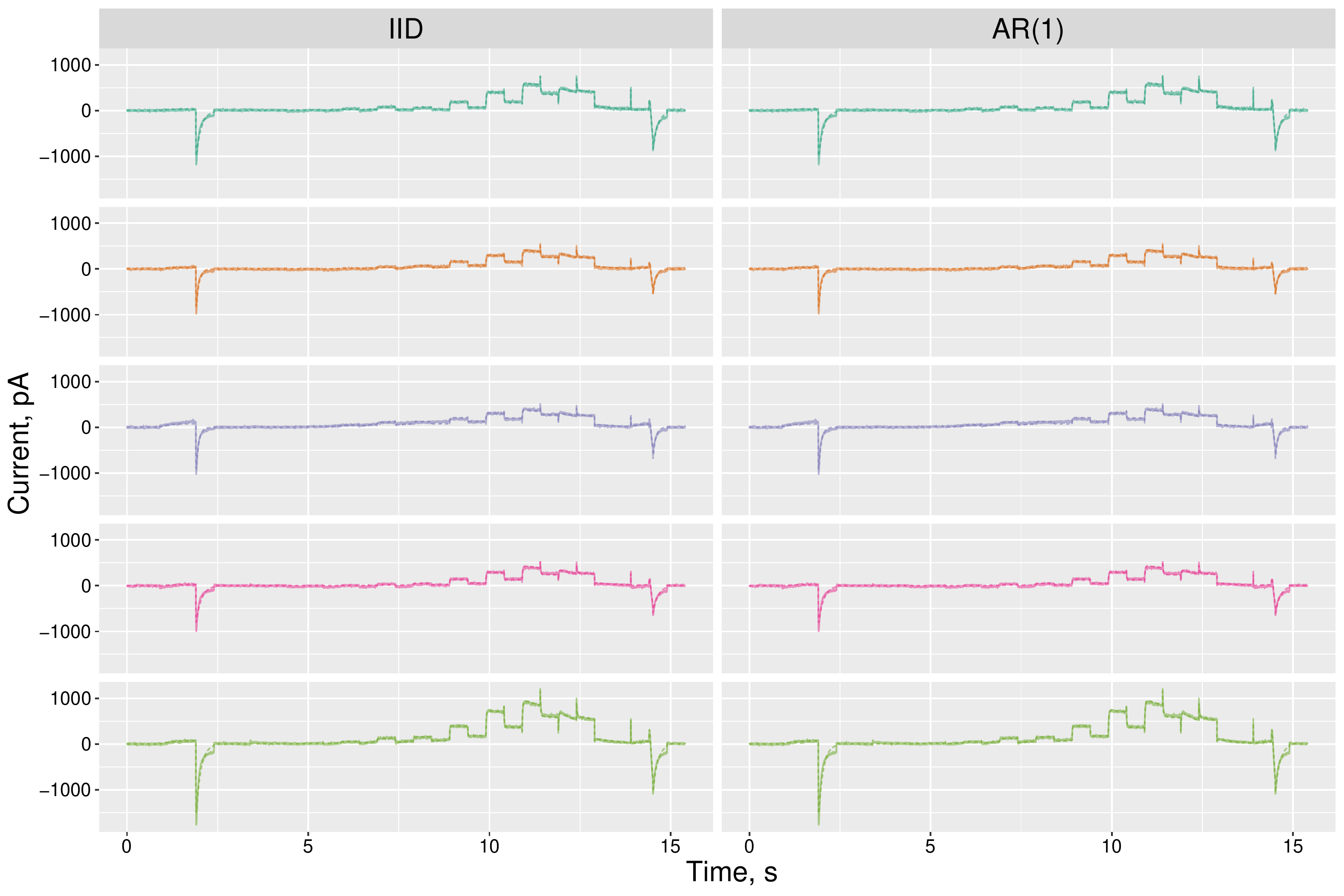}}
	\caption{\textbf{hERG model: posterior predictions.} The dashed lines show posterior median predictions and the solid lines show experimental data (here, it is difficult to visually discern a difference between model fits and data). Each row shows the results for each of the five cells: colours indicate the cell and correspond with those shown in Fig. \ref{fig:herg_posteriors}. Columns show the results when assuming either IID Gaussian noise or AR(1) noise.}
	\label{fig:herg_posterior_predictive}
\end{figure}

In Fig. \ref{fig:herg_posteriors}, we compare the posterior distributions for the model parameters obtained across both noise models. For some parameters: $g_{Kr}, p_1, p_2$ and $p_6$, the estimates were similar across both the IID and AR(1) models; for others: $p_3, p_4, p_5, p_7$ and $p_8$, there were often substantial differences. Despite these differences in parameter values, the IID and AR(1) models appeared visually to fit the data equally well (Fig. \ref{fig:herg_posterior_predictive}). The extent to which the estimates differed also depended on the cell in question, with the cells shown in pink and dark green generally showing greater discrepancies.

\begin{figure}[ht]
	\centerline{\includegraphics[width=1\textwidth]{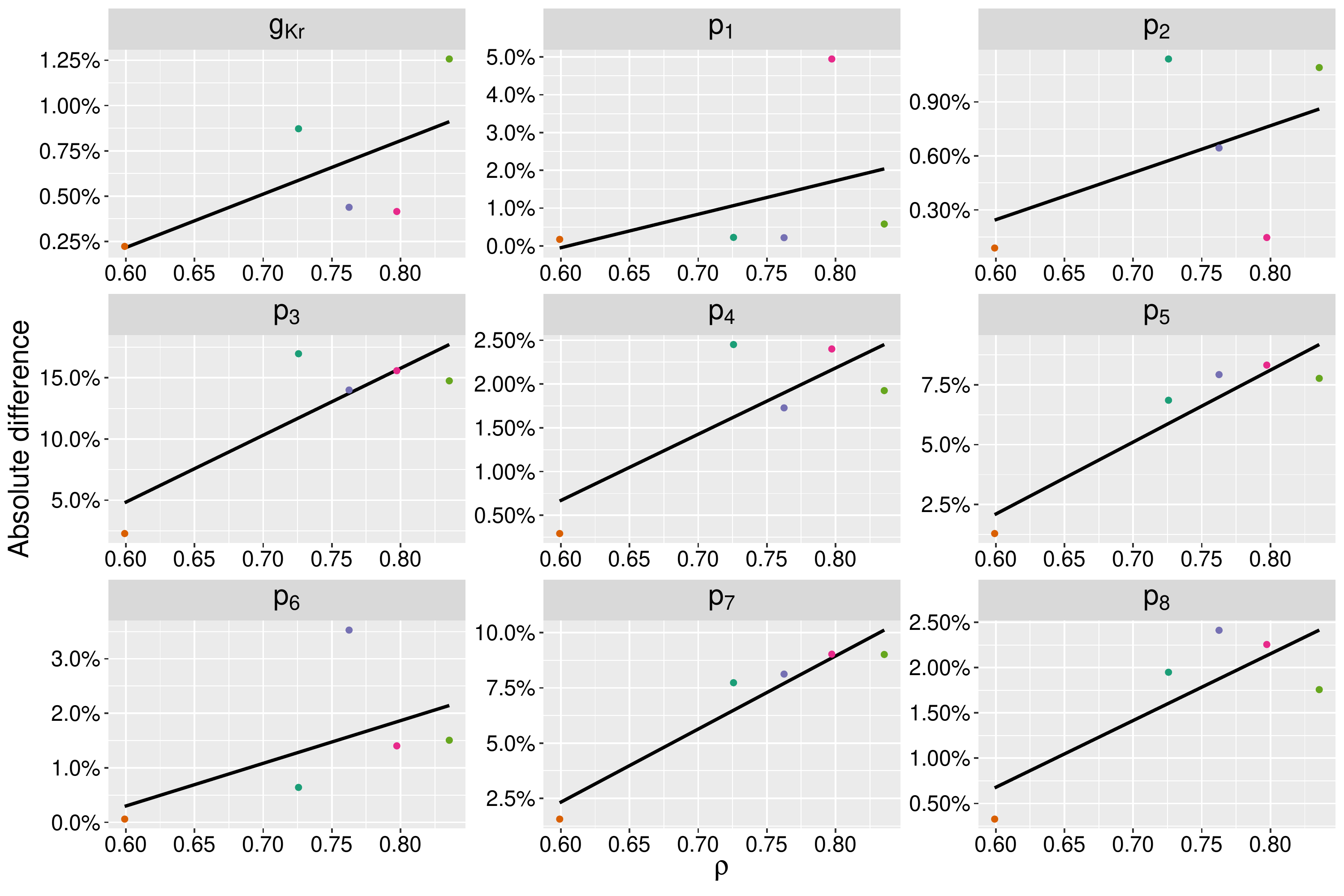}}
	\caption{\textbf{hERG model: point estimate differences.} The horizontal axis shows the point estimate of $\rho$ in eq. \eqref{eq:ar1} estimated for each cell. The vertical axis shows the absolute percentage difference between the IID estimates and the AR(1) ones (the denominator is the IID estimate). Colours indicate the experimental replicate (i.e. the cell on which experiments were performed) and correspond with those shown in Fig. \ref{fig:herg_posteriors}. The lines show linear ordinary least-squares regression fits done separately for each parameter.}
	\label{fig:herg_differences}
\end{figure}

To further investigate the cause of these discrepancies, in Fig. \ref{fig:herg_differences}, we plot the posterior median $\rho$ value from the AR(1) model versus the absolute percentage difference between the IID and AR(1) models. We also plot the best fit lines (in black) from linear regressions of the absolute difference on $\rho$ for each parameter. Across all parameters, these indicate that as the magnitude of estimated error autocorrelation increased, there were greater differences between the IID and AR(1) model estimates.

\begin{figure}[ht]
	\centerline{\includegraphics[width=1\textwidth]{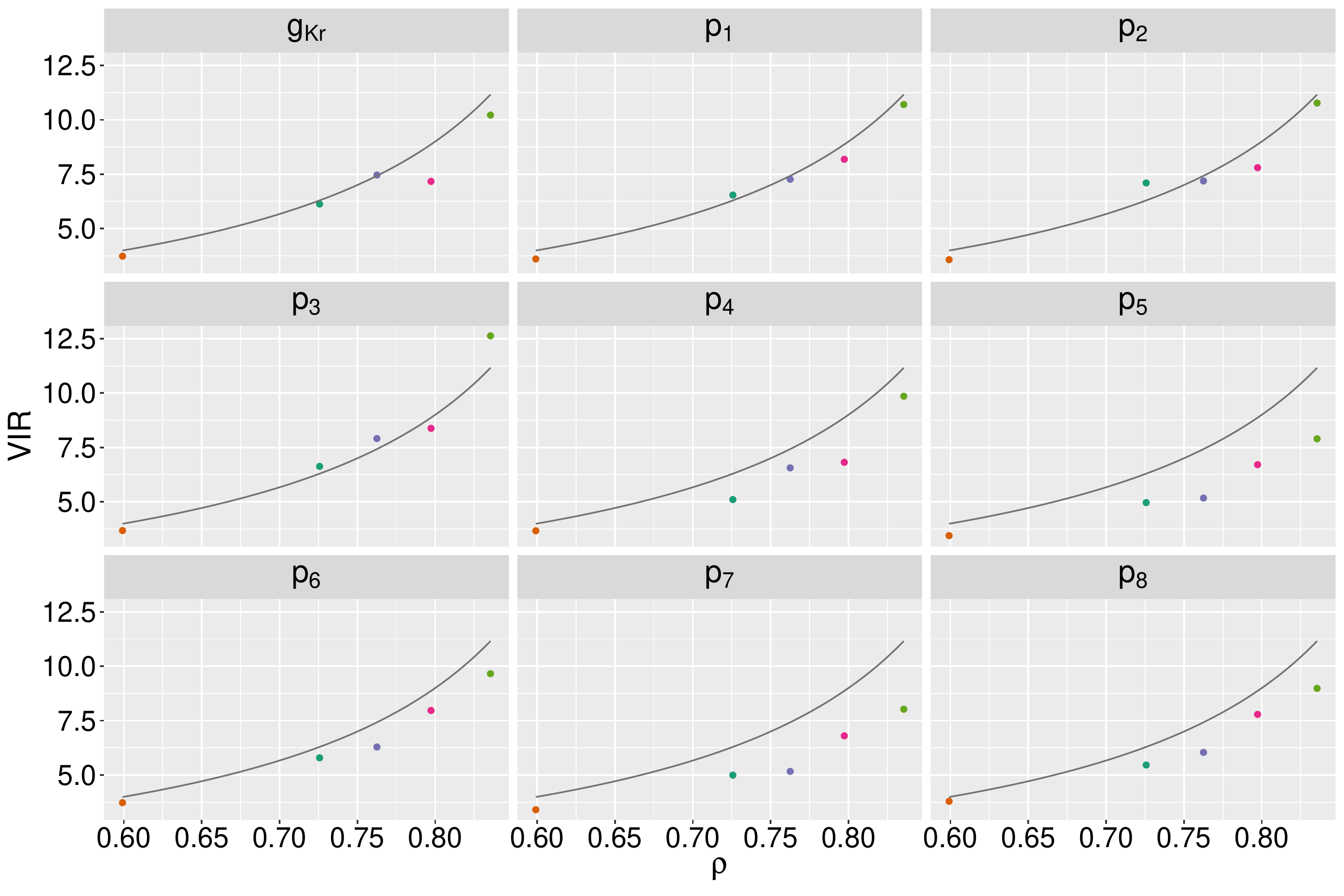}}
	\caption{\textbf{hERG model: VIRs.} The horizontal axis shows the $\rho$ posterior median values estimated from the AR(1) noise model; the vertical axis shows the estimated VIR for each parameter. Colours indicate the experimental replicate (i.e. the cell on which experiments were performed) and correspond with those shown in Fig. \ref{fig:herg_posteriors}. The line shows the theoretical VIR for a linear model described by eq. \eqref{eq:vif_constant}.}
	\label{fig:herg_vir}
\end{figure}

Finally, we estimate VIRs for each parameter across all cells in the system by taking the ratio of the AR(1) posterior variance to the IID equivalent. In Fig. \ref{fig:herg_vir}, we plot these versus the estimated $\rho$ value for each cell. In all cases, as $\rho$ increased, the VIRs followed suit. In the same plot, we also overlay the theoretical VIR given by eq. \eqref{eq:vif_constant} for a linear model, since the nonlinear case is not straightforward to calculate for this model. Whilst the hERG model is nonlinear and the true noise process is unknown, in many cases, the theoretical VIR provided a reasonable guide as to how the variance increased with $\rho$.

\subsection{Electrochemistry model}\label{sec:electrochemistry}
We next apply our methodology to a system in electrochemistry: unlike the previous examples, the model here is a partial differential equation, although yielding a single output -- a current -- which we fit to data. Since none of the theory derived in \S\ref{sec:methods_nonliear_ode} assumes a particular form of the function, the results are not bespoke for ODEs. And, because the PDE has only a single output time series, we use the same statistical framework as for our other examples. Further details are provided in \S\ref{sec:appendix_electrochemistry}.

In this example, we observed current time series, $\{\tilde{I}_{\tiny{tot}}(t)\}$ resulting from a laboratory experiment. We assumed that $\tilde{I}_{\tiny{tot}}(t) = I_{\tiny{tot}}(t) + \epsilon(t)$, where $\epsilon(t)$ is an error process. We fixed a series of parameters in the model to experimentally determined values as given in Table \ref{tab:electrochem_experimental_parameters}. On the remaining six parameters, we placed uniform priors as given in Table \ref{tab:electrochem_inference_parameters}.

\begin{figure}[ht]
	\centerline{\includegraphics[width=0.8\textwidth]{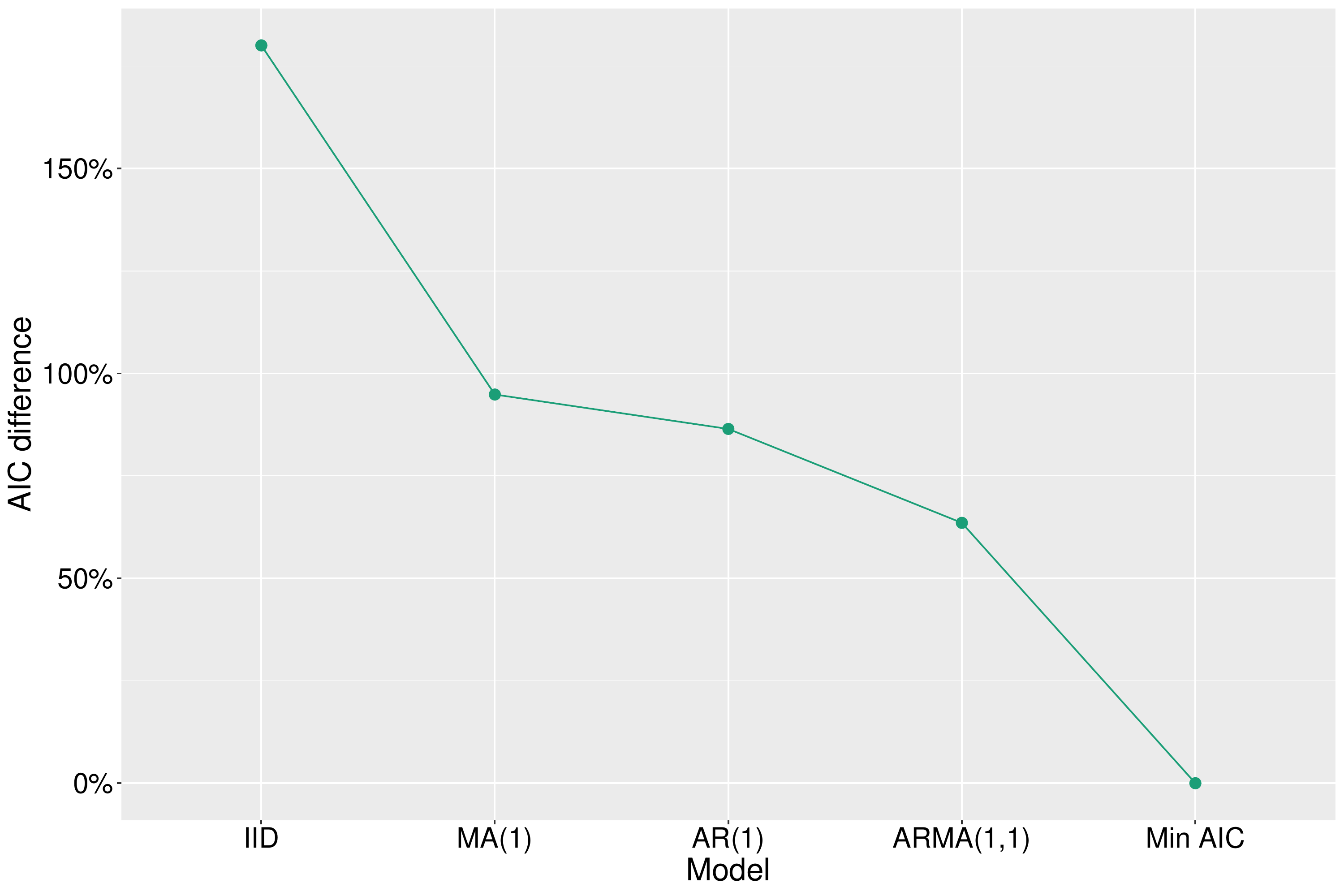}}
	\caption{\textbf{Electrochemistry model: AICs.} The vertical axis shows the percentage difference between the AIC of each model (indicated on horizontal axis) compared to the best fitting model (``Min AIC'').}
	\label{fig:electro_arima_aics}
\end{figure}

To assess the level of autocorrelation in the error process, we follow the approach outlined in \S\ref{sec:diagnosis}. In particular, we assumed that the noise process is IID Gaussian and used an optimiser, CMA-ES \cite{hansen2016cma} (as implemented in PINTS \cite{Clerx2019Pints}), to determine maximum likelihood estimates of the parameter values and to obtain a residual series. We then compared the fit of various ARIMA models to these residuals: in Fig. \ref{fig:electro_arima_aics}, we compare the AICs from IID, MA(1), AR(1) and ARMA(1,1) models to the one which minimised this criterion: an ARIMA(4, 1, 4) model. This shows that the IID Gaussian model is substantially bettered by models incorporating autocorrelation in the error series.

As part of this process, we also fitted to the residual series using various types of state-space models. To do this fitting, we relied on the \textit{Statsmodels} Python package \cite{seabold2010statsmodels}. The state-space models we tried included a \textit{local level} model, a \textit{random walk with drift} model and a \textit{random trend} model: all of these had substantially worse fits as determined by AIC compared to the ARIMA processes. Because of this, we did not go ahead with full Bayesian inference for these model types.

We next attempted to fit the electrochemistry model assuming AR(1), ARMA(1,1) and the ARIMA(4,1,4) error processes in a Bayesian model; we also fitted the model using a IID Gaussian error process for comparison. The models were fitted using the Haario-Bardenet adaptive-covariance MCMC algorithm in PINTS \cite{Clerx2019Pints}. Uniform priors were set on all fitted parameters as described in \cite{robinson2018separating}. The Markov chains were initialised to the MAP points found using the CMA-ES optimisation algorithm. Three chains were run using 10,000 samples, the first 3,000 of which were discarded as warm-up. Convergence was diagnosed via $\hat{R}<1.1$. We were unable to obtain Markov chain convergence for the ARIMA(4,1,4) model: we speculate that this was because the additional number of parameters of this model caused the inferred errors themselves to become unidentified.

\begin{figure}[ht]
	\centerline{\includegraphics[width=1\textwidth]{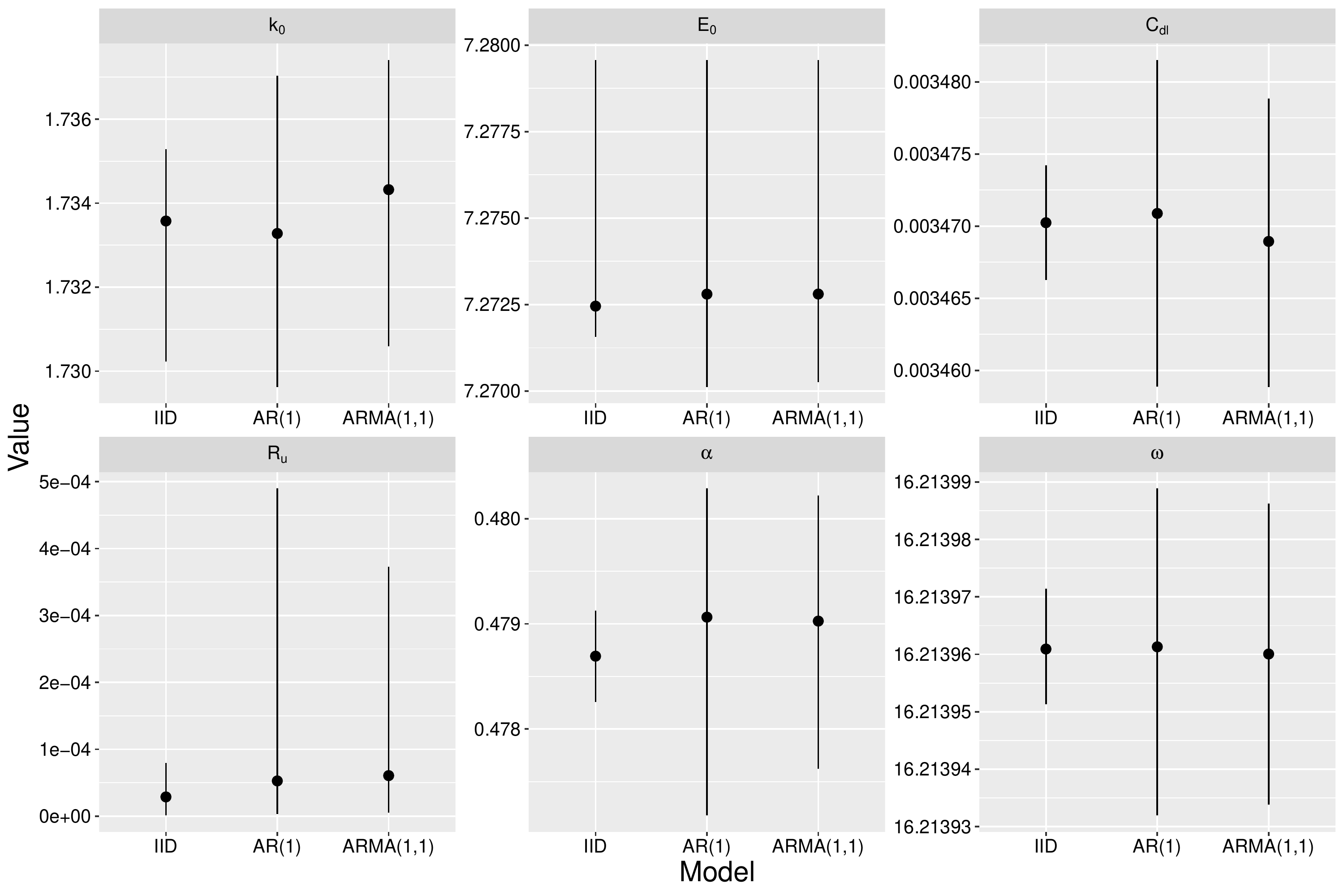}}
	\caption{\textbf{Electrochemistry model: posterior distributions.} Each panel displays posterior summaries for the IID, AR(1) and ARMA(1,1) models for each inferred parameter. The points show the posterior median, and the upper and lower samples show the 97.5\% and 2.5\% posterior quantiles.}
	\label{fig:electro_posteriors}
\end{figure}

In Fig. \ref{fig:electro_posteriors}, we show the estimated posteriors for the IID, AR(1) and ARMA(1,1) models. In this figure, the panels show posterior summaries for each parameter across the three models. Across all parameters, the AR(1) and ARMA(1,1) models had increased uncertainty relative to the IID model. This was most notable for the uncompensated resistance $R_u$, where the two models with autocorrelated errors produced distributions with longer tails. Additionally, the median point estimates of parameters varied across the three models (again, most notably for $R_u$).

\section{Discussion}
This work highlights how mischaracterising the measurement process for ODE models can have marked consequences for inference. Our results indicate that failing to account for measurement-induced autocorrelation in errors results in overconfident estimates of parameter values, with the degree of overconfidence depending on the magnitude and type of stochastic process governing measurements. Using real data collected from experiments in cardiac electrophysiology and electrochemistry, we fit models assuming independent errors and obtain residual series that bear the hallmarks of autocorrelated errors. When these models were refitted assuming autocorrelated noise processes, we obtained considerably wider parameter bounds than when specifying independent noise. Whether this is a more general phenomena is unclear, but our results indicate that choice of measurement process can substantially affect inference. So choice of measurement process needs to be done with due care, and the types of diagnostic plots we use here can help to guide this process.

Misspecification of the ODE model can also generate autocorrelated errors, but its impact on inferences is likely different. When an ODE model is misspecified, parameter estimates (if these same parameters span both the correct and misspecified models) may display bias due to parameter compensation \cite{kennedy2001bayesian,brynjarsdottir2014learning}. Error autocorrelation due to model misspecification could, in some cases, be modelled using some of the noise processes we describe here. Whether they should be, however, is less clear. It is possible that the two example systems we investigated did involve misspecified models, and part of the observed autocorrelation was due to this. We found that, by accounting for an autocorrelated error process, the uncertainty in the estimates was generally wider and, in some cases, the point estimates deviated considerably from the null IID Gaussian model. Because these are real life models, however, it is not straightforward to determine whether using an autocorrelated error model led to improved estimates. Future work, using toy models with known misspecifications and autocorrelated measurement processes, could shed light on how to best to account for both issues.

In this work, we considered a range of noise processes including ARMA and STS models. For our applied examples, sometimes complex autocorrelation structures were found to best fit the error variation, and it is questionnable whether  measurement processes could have generated these errors. Additionally, in some circumstances, the imposition of such measurement processes rendered the system practically unidentified, an issue with error processes which has long been recognised \cite{kleibergen1997bayesian}. So how should an appropriate noise model be chosen? A noise process is itself a model, albeit a statistical one. Like other elements of the system, it should be understandable: if it is overly complex, the noise process is more likely to overfit current data resulting in poor generalisation of the overall model. By contrast, when assuming independent noise, this can also often produce parameter sets that are more likely to overfit current data. We, hence, argue that using a low order ARMA model or a relatively simple STS noise model is preferable in many circumstances by helping to guard against some of the larger effects of measurement model misspecification. We do not make rigid specifications as to the limiting complexity of these processes that are used, but believe a reasonable litmus test is, ``Could I convince a colleague that this noise process represents the actual measurement process?''. If the measurement process is well understood and arguments can be made for complex measurement processes, then this reasoning should be explicitly stated.

More mechanistic models of the measurement process may also lead to clearer understanding of the underlying biological processes. A recent study modelled the measurement process of patch-clamp experiments, accounting for series resistance, membrane and pipette capacitance, voltage offsets, imperfect compensations made by the amplifier, and leak currents \cite{lei_accounting_2019}. In explaining inter-cell variation through imperfections in measurement, this produced a more parsimonious explanation of the data than when assuming cell-specific ion current kinetics. Another study from parasitology examined laboratory experiments, where mosquitoes are infected with malaria parasites through membrane feeding assays \cite{stopard2021estimating}. By considering the measurement processes leading to observations -- that experiments consist of mosquitoes being randomly sampled from a wider pool of specimens and each dissection representing an individual snapshot of the parasite dynamics -- this resulted in novel estimates of key parameters in epidemiology.

Here, we considered only noise processes which had a fixed form over time, meaning our analysis does not consider either temporal or output-linked heteroscedasticity. Nor do the noise models we consider allow the autocorrelation structure itself to change with time. Recent work in related systems has shown that time-varying noise processes may provide a better representation, where, typically, throughout a time trace of an output variable, there are some regions of low autocorrelation and low variation punctuated by high autocorrelation / high uncertainty regions \cite{creswell2020using}. The general noise processes used to handle these temporal patterns are likely to be non-parametric and less amenable to direct analysis than the processes we consider here. But, our analytical results may nonetheless provide an approximate guide as to the impact on parameter inference of modelling noise using non-IID processes. We also did not consider measurement of multiple states of a system and the possible correlations across these, which, intuitively, should reduce the information content of observations. It has been empirically demonstrated that choosing so-called \textit{robust} error models, such as the Student-t and Huber distributions can lead to better estimates \cite{maier2017robust}, and it is possible that the techniques we use here could produce useful analytical results when applied to those situations.

In systems where the state is measured repeatedly over short time intervals, such as those in electrochemistry, cardiac physiology and neuroscience, experimental limitations may mean that the assumption of independent measurements is suspect. In these types of systems, it may thus be better to assume an autocorrelated measurement model by default to mitigate against the risk of unrealistically precise estimates. As experimental methods are developed to allow collection of data at increasingly finer gradations, however, accounting for measurement imperfections will likely be increasingly important when performing inference.

\bibliography{Bayes}

\begin{thebibliography}{10}

\bibitem{anderson1992infectious}
RM~Anderson and RM~May.
\newblock {\em Infectious diseases of humans: dynamics and control}.
\newblock Oxford University Press, 1992.

\bibitem{murray2007mathematical}
JD~Murray.
\newblock {\em Mathematical biology: I. An Introduction (interdisciplinary
  applied mathematics)(Pt. 1)}.
\newblock New York, Springer, 2007.

\bibitem{hodgkin1952quantitative}
Alan~L Hodgkin and Andrew~F Huxley.
\newblock A quantitative description of membrane current and its application to
  conduction and excitation in nerve.
\newblock {\em The Journal of physiology}, 117(4):500, 1952.

\bibitem{ashyraliyev2009systems}
M~Ashyraliyev, Y~Fomekong-Nanfack, JA~Kaandorp, and JG~Blom.
\newblock Systems biology: parameter estimation for biochemical models.
\newblock {\em The FEBS journal}, 276(4):886--902, 2009.

\bibitem{mendes1998non}
P~Mendes and D~Kell.
\newblock Non-linear optimization of biochemical pathways: applications to
  metabolic engineering and parameter estimation.
\newblock {\em Bioinformatics (Oxford, England)}, 14(10):869--883, 1998.

\bibitem{gabor2015robust}
A~G{\'a}bor and JR~Banga.
\newblock Robust and efficient parameter estimation in dynamic models of
  biological systems.
\newblock {\em BMC Systems Biology}, 9(1):1--25, 2015.

\bibitem{vanlier2013parameter}
J~Vanlier, CA~Tiemann, PAJ Hilbers, and NAW Van~Riel.
\newblock Parameter uncertainty in biochemical models described by ordinary
  differential equations.
\newblock {\em Mathematical Biosciences}, 246(2):305--314, 2013.

\bibitem{villaverde2019benchmarking}
AF~Villaverde, F~Fr{\"o}hlich, D~Weindl, J~Hasenauer, and JR~Banga.
\newblock Benchmarking optimization methods for parameter estimation in large
  kinetic models.
\newblock {\em Bioinformatics}, 35(5):830--838, 2019.

\bibitem{girolami2008bayesian}
M~Girolami.
\newblock Bayesian inference for differential equations.
\newblock {\em Theoretical Computer Science}, 408(1):4--16, 2008.

\bibitem{jaynes2003probability}
ET~Jaynes.
\newblock {\em Probability theory: The logic of science}.
\newblock Cambridge University Press, 2003.

\bibitem{maier2017robust}
C~Maier, C~Loos, and J~Hasenauer.
\newblock Robust parameter estimation for dynamical systems from
  outlier-corrupted data.
\newblock {\em Bioinformatics}, 33(5):718--725, 2017.

\bibitem{simoen2013prediction}
E~Simoen, C~Papadimitriou, and G~Lombaert.
\newblock On prediction error correlation in bayesian model updating.
\newblock {\em Journal of Sound and Vibration}, 332(18):4136--4152, 2013.

\bibitem{kennedy2001bayesian}
MC~Kennedy and A~O'Hagan.
\newblock Bayesian calibration of computer models.
\newblock {\em Journal of the Royal Statistical Society: Series B (Statistical
  Methodology)}, 63(3):425--464, 2001.

\bibitem{brynjarsdottir2014learning}
J~Brynjarsd{\'o}ttir and A~O'Hagan.
\newblock Learning about physical parameters: The importance of model
  discrepancy.
\newblock {\em Inverse Problems}, 30(11):114007, 2014.

\bibitem{lyddon2018nonparametric}
S~Lyddon, S~Walker, and C~Holmes.
\newblock Nonparametric learning from bayesian models with randomized objective
  functions.
\newblock In {\em Advances in Neural Information Processing Systems}, pages
  2071--2081, 2018.

\bibitem{lei2020considering}
CL~Lei, S~Ghosh, DG~Whittaker, Y~Aboelkassem, KA~Beattie, CD~Cantwell,
  T~Delhaas, C~Houston, GM~Novaes, and AV~Panfilov.
\newblock Considering discrepancy when calibrating a mechanistic
  electrophysiology model.
\newblock {\em Philosophical Transactions of the Royal Society A},
  378(2173):20190349, 2020.

\bibitem{wooldridge2015introductory}
JM~Wooldridge.
\newblock {\em Introductory econometrics: A modern approach}.
\newblock Nelson Education, 2015.

\bibitem{harvey1990forecasting}
AC~Harvey.
\newblock {\em Forecasting, structural time series models and the {Kalman}
  filter}.
\newblock Cambridge University Press, 1990.

\bibitem{box2015time}
GEP Box, GM~Jenkins, GC~Reinsel, and GM~Ljung.
\newblock {\em Time series analysis: forecasting and control}.
\newblock John Wiley \& Sons, 2015.

\bibitem{hamilton1994time}
JD~Hamilton.
\newblock {\em Time series analysis}, volume~2.
\newblock Princeton New Jersey, 1994.

\bibitem{carpenter2016stan}
B~Carpenter, A~Gelman, M~Hoffman, D~Lee, B~Goodrich, M~Betancourt, MA~Brubaker,
  J~Guo, P~Li, and A~Riddell.
\newblock Stan: A probabilistic programming language.
\newblock {\em J Stat Softw}, 2016.

\bibitem{hoffman2014no}
MD~Hoffman and A~Gelman.
\newblock The {No-U-turn} sampler: adaptively setting path lengths in
  {H}amiltonian {M}onte {C}arlo.
\newblock {\em Journal of Machine Learning Research}, 15(1):1593--1623, 2014.

\bibitem{gelman1992inference}
A~Gelman and DB~Rubin.
\newblock Inference from iterative simulation using multiple sequences.
\newblock {\em Statistical Science}, pages 457--472, 1992.

\bibitem{lei_rapid_2019-1}
CL~Lei, M~Clerx, DJ~Gavaghan, L~Polonchuk, GR~Mirams, and K~Wang.
\newblock Rapid characterisation of {hERG} channel kinetics {I}: using an
  automated high-throughput system.
\newblock {\em Biophysical Journal}, 117:2438--2454, 2019.

\bibitem{lei_rapid_2019-2}
CL~Lei, M~Clerx, KA~Beattie, D~Melgari, JC~Hancox, DJ~Gavaghan, L~Polonchuk,
  K~Wang, and GR~Mirams.
\newblock Rapid characterisation of {hERG} channel kinetics {II}: temperature
  dependence.
\newblock {\em Biophysical Journal}, 117:2455--2470, 2019.

\bibitem{hansen2016cma}
N~Hansen.
\newblock The {CMA} evolution strategy: A tutorial.
\newblock {\em arXiv preprint arXiv:1604.00772}, 2016.

\bibitem{Clerx2019Pints}
M~Clerx, M~Robinson, B~Lambert, CL~Lei, S~Ghosh, GR~Mirams, and DJ~Gavaghan.
\newblock Probabilistic inference on noisy time series ({PINTS}).
\newblock {\em Journal of Open Research Software}, 7(1):23, 2019.

\bibitem{lei_accounting_2019}
CL~Lei, M~Clerx, DG~Whittaker, DJ~Gavaghan, TP~de~Boer, and GR~Mirams.
\newblock Accounting for variability in ion current recordings using a
  mathematical model of artefacts in voltage-clamp experiments.
\newblock {\em Philosophical Transactions of The Royal Society A},
  378(2173):20190348, 2020.

\bibitem{akaike1974new}
H~Akaike.
\newblock A new look at the statistical model identification.
\newblock {\em IEEE transactions on automatic control}, 19(6):716--723, 1974.

\bibitem{jasra2007population}
A~Jasra, DA~Stephens, and CC~Holmes.
\newblock On population-based simulation for static inference.
\newblock {\em Statistics and Computing}, 17(3):263--279, 2007.

\bibitem{kleibergen1997bayesian}
F~Kleibergen and H~Hoek.
\newblock Bayesian analysis of {ARMA}models using noninformative priors.
\newblock {\em Tinbergen Institute discussion paper}, 1997.

\bibitem{seabold2010statsmodels}
S~Seabold and J~Perktold.
\newblock Statsmodels: Econometric and statistical modeling with python.
\newblock {\em Proceedings of the 9th Python in Science Conference}, 57:61,
  2010.

\bibitem{robinson2018separating}
M~Robinson, AN~Simonov, J~Zhang, AM~Bond, and D~Gavaghan.
\newblock Separating the effects of experimental noise from inherent system
  variability in voltammetry: The \ch{[{F}e ({CN})_6] ^{3-/4-}}process.
\newblock {\em Analytical Chemistry}, 91(3):1944--1953, 2018.

\bibitem{stopard2021estimating}
IJ~Stopard, TS~Churcher, and B~Lambert.
\newblock Estimating the extrinsic incubation period of malaria using a
  mechanistic model of sporogony.
\newblock {\em PLoS computational biology}, 17(2):e1008658, 2021.

\bibitem{creswell2020using}
R~Creswell, B~Lambert, CL~Lei, M~Robinson, and D~Gavaghan.
\newblock Using flexible noise models to avoid noise model misspecification in
  inference of differential equation time series models.
\newblock {\em arXiv preprint arXiv:2011.04854}, 2020.

\bibitem{zivot2006state}
E~Zivot.
\newblock State space models and the {Kalman} filter, 2006.
\newblock [Online; accessed 26-Jun-2020].

\bibitem{wiki2020normal}
{Wikipedia contributors}.
\newblock Multivariate normal distribution--- {W}ikipedia{,} the free
  encyclopedia, 2020.
\newblock [Online; accessed 19-Jun-2020].

\bibitem{sher2004resistance}
AA~Sher, AM~Bond, DJ~Gavaghan, K~Harriman, SW~Feldberg, NW~Duffy, SX~Guo, and
  J~Zhang.
\newblock Resistance, capacitance, and electrode kinetic effects in
  fourier-transformed large-amplitude sinusoidal voltammetry: Emergence of
  powerful and intuitively obvious tools for recognition of patterns of
  behavior.
\newblock {\em Analytical Chemistry}, 76(21):6214--6228, 2004.

\bibitem{morris2015theoretical}
GP~Morris, RE~Baker, K~Gillow, JJ~Davis, DJ~Gavaghan, and AM~Bond.
\newblock Theoretical analysis of the relative significance of thermodynamic
  and kinetic dispersion in the dc and ac voltammetry of surface-confined
  molecules.
\newblock {\em Langmuir}, 31(17):4996--5004, 2015.

\end{thebibliography}
\bibliographystyle{unsrt}
\clearpage

\beginsupplement
\section{Nonlinear differential equations}
\subsection{Nonlinear differential equation: single parameter}\label{sec:methods_ode}
We now consider a model of the form,
\begin{equation}\label{eq:nonlinear_system2}
x(t) = f(t;\theta) + \epsilon(t),
\end{equation}
where, for example, $f(t;\theta)$ is the solution of a nonlinear ODE (or a function of the solution of such an ODE) with univariate parameter $\theta$. As before, the true error process is AR(1) as given by eq. \eqref{eq:ar1}. The IID Gaussian random variable is now given by,
\begin{equation}
\nu(t) = x(t) - \rho x(t-1) - (f(t;\theta) - \rho f(t-1;\theta)).
\end{equation}
The second derivative of the true model log-likelihood is, hence, given by,
\begin{equation}\label{eq:nonlinear_information}
\frac{\partial^2\mathcal{L}}{\partial\theta^2} = -\frac{1}{\sigma^2}\left\{\sum_{t=1}^{T}\left(\frac{\partial f}{\partial \theta}\Bigr|_{t,\theta} - \rho \frac{\partial f}{\partial \theta}\Bigr|_{t-1,\theta}\right)^2 + \sum_{t=1}^{T} \nu(t) \left(\frac{\partial^2 f}{\partial \theta^2}\Bigr|_{t,\theta} - \rho \frac{\partial^2 f}{\partial \theta^2}\Bigr|_{t-1,\theta}\right)\right\}.
\end{equation}
On expectation, the second term in eq. \eqref{eq:nonlinear_information} becomes zero, yielding the following expression for the diagonal element corresponding to $\theta$ in the information matrix,
\begin{equation}\label{eq:nonlinear_information1}
\mathcal{I}_{\theta,\theta}=\frac{1}{\sigma^2} \sum_{t=1}^{T}\left(\frac{\partial f}{\partial \theta}\Bigr|_{t,\theta} - \rho \frac{\partial f}{\partial \theta}\Bigr|_{t-1,\theta}\right)^2,
\end{equation}
with the CRLB given by the inverse of this quantity, 
\begin{equation}\label{eq:nonlinear_crlb}
\text{var}(\hat{\theta}) = \sigma^2 / \sum_{t=1}^{T}\left(\frac{\partial f}{\partial \theta}\Bigr|_{t,\theta} - \rho \frac{\partial f}{\partial \theta}\Bigr|_{t-1,\theta}\right)^2.
\end{equation}
The equivalent CRLB for the model assuming independent errors is obtained by substituting $\rho=0$ into eq. \eqref{eq:nonlinear_crlb} and replacing $\sigma^2$ with ${\sigma'^2= \sigma^2/(1-\rho^2)}$ to account for the different variance estimated by the independent error model (as was done in \S\ref{sec:methods_constant}). The variance inflation ratio for a nonlinear model is then given by the ratio of these two CRLBs:
\begin{equation}\label{eq:vif_nonlinear1}
\text{VIR}(\rho) =  (1-\rho^2)\sum_{t=1}^{T}\left(\frac{\partial f}{\partial \theta}\Bigr|_{t,\theta}\right)^2 / \sum_{t=1}^{T}\left(\frac{\partial f}{\partial \theta}\Bigr|_{t,\theta} - \rho \frac{\partial f}{\partial \theta}\Bigr|_{t-1,\theta}\right)^2.
\end{equation}

\subsection{Nonlinear differential equation: multiple parameters}\label{sec:methods_ode_multiple}
We now consider a univariate model of the form,
\begin{equation}\label{eq:nonlinear_system_m}
x(t) = f(t;\theta) + \epsilon(t),
\end{equation}
where $\theta\in \mathbb{R}^m$, with the true error process given by eq. \eqref{eq:ar1}. Defining $f_{\theta_i}(t):=\frac{\partial f}{\partial \theta_i}\Bigr|_{t-1,\theta_i}$ and:
\begin{equation}
g(\theta_i, \theta_j, \rho):=\sum_{t=1}^T (f_{\theta_i}(t) - \rho f_{\theta_i}(t-1)) (f_{\theta_j}(t) - \rho f_{\theta_j}(t-1)),    
\end{equation}
we can write the general information matrix (excluding the terms for $\sigma^2$, since it is uncorrelated with the model parameters):
\begin{equation}
    \mathcal{I}(\rho) = \frac{1}{\sigma^2}\begin{bmatrix}
    g(\theta_1, \theta_1, \rho) & g(\theta_1, \theta_2,\rho) & \dots & g(\theta_1, \theta_m,\rho)\\
    g(\theta_2, \theta_1,\rho) & g(\theta_2, \theta_2,\rho) & \dots & g(\theta_2, \theta_m,\rho)\\
     \vdots & \vdots & & \vdots\\
     g(\theta_m, \theta_1,\rho) & g(\theta_m, \theta_2,\rho) & \dots & g(\theta_m, \theta_m,\rho)\\
    \end{bmatrix}.
\end{equation}
The CRLB is given by $A(\rho)=\mathcal{I}(\rho)^{-1}$ with the $i$th diagonal elements giving the asymptotic variance in the estimate of $\theta_i$. This leads to an expression for the VIR for parameter $\theta_i$ given by:
\begin{equation}\label{eq:vir_theoretical_exact}
    \text{VIR}(\theta_i) = (1 - \rho^2) \frac{A_{i,i}(\rho)}{A_{i,i}(0)}.
\end{equation}
\subsection{Nonlinear differential equations: unknown $\sigma$ parameter}\label{sec:methods_ode_sigma}
Considering a nonlinear model of the same form as in section \ref{sec:methods_ode_multiple}, we now suppose that the parameter, $\sigma>0$, characterising the variance of the error process is unknown. To determine the VIR in this circumstance, we differentiate the log-likelihood with respect to $\sigma$:
\begin{equation}
    \frac{\partial \mathcal{L}}{\partial \sigma} = -\frac{T}{\sigma} + \frac{1}{\sigma^3} \sum_{t=1}^T \nu(t)^2.
\end{equation}
Differentiating the above with respect to $\theta_i$, we obtain:
\begin{equation}
    \mathbb{E}\left[-\frac{\partial^2 \mathcal{L}}{\partial \sigma\theta_i}\right] = -\frac{2}{\sigma^3} \sum_{t=1}^T \frac{\partial\nu(t)}{\partial\theta_i}\mathbb{E}[\nu(t)]=0,
\end{equation}
because $\frac{\partial\nu(t)}{\partial\theta_i}$ is deterministic, and $\mathbb{E}[\nu(t)]=0$. The FIM for model parameters is thus unaffected by uncertainty around $\sigma$.

\subsection{Nonlinear differential equations: unknown AR(1) parameter}\label{sec:methods_ode_rho}
Considering a nonlinear model of the same form as in section \ref{sec:methods_ode_multiple}, we now suppose that the parameter, $\rho$, characterising the degree of autocorrelation in the AR(1) errors. To determine the VIR in this situation, we consider the derivative of the log-likelihood with respect to $\rho$:
\begin{equation}
    \frac{\partial \mathcal{L}}{\partial \rho} = -\frac{1}{\sigma^2} \sum_{t=1}^T \nu(t) \frac{\partial \nu(t)}{\partial \rho}
\end{equation}
where $\nu(t)=x(t) - \rho x(t-1) - (f(t; \theta) - \rho f(t-1;\theta))$, so $\frac{\partial \nu(t)}{\partial \rho} = f(t;\theta)-x(t)$. To determine the FIM, we next consider the off-diagonal second-order partial derivatives of the log-likelihood:
\begin{equation}
    \frac{\partial^2 \mathcal{L}}{\partial \rho\theta_i} = -\frac{1}{\sigma^2} \sum_{t=1}^T \left\{\frac{\partial \nu(t)}{\partial \rho}\frac{\partial \nu(t)}{\partial \theta_i} + \nu(t)\frac{\partial^2 \nu(t)}{\partial \rho\partial \theta_i} \right\}
\end{equation}
where $\frac{\partial \nu(t)}{\partial \theta_i} = -(f_{\theta_i}(t) - \rho f_{\theta_i}(t-1))$. The corresponding elements of the FIM are given by:
\begin{equation}
    \mathbb{E}\left[-\frac{\partial^2 \mathcal{L}}{\partial \rho\theta_i}\right] = \frac{1}{\sigma^2} \sum_{t=1}^T \left\{ -(f_{\theta_i}(t) - \rho f_{\theta_i}(t-1))(f(t-1;\theta)-\mathbb{E}\left[x(t-1))\right] + f_{\theta_i}(t-1)\mathbb{E}\left[\nu(t)\right]\right\} = 0.
\end{equation}
Since the off-diagonal terms in the FIM between model parameters, $\theta_i$, and $\rho$ are zero, uncertainty about $\rho$ should not influence VIRs obtained on other parameters.

\subsection{Nonlinear differential equation: unknown initial state}\label{sec:methods_ode_initial state}
In many initial value problems, the initial state of the system, $x_0:=x(0)$, is unknown and must also be inferred from data. Considering a nonlinear model of the same form as in section \ref{sec:methods_ode_multiple}, we now determine the VIRs when the initial state is also unknown.

There are two distinct contributions to the FIM due to $x_0$: one due to the dependence of the solution, $f(.)$, on the initial conditions; the other because the initial condition appears in the first term of the log-likelihood summation:
\begin{equation}
    \mathcal{L} \sim -\frac{1}{2\sigma^2} \left(x(1) - \rho x(0) - (f(1;\theta) - \rho f(0;\theta)\right)^2 - \frac{1}{2\sigma^2}\sum_{t=2}^T \left(x(t) - \rho x(t-1) - (f(t;\theta) - \rho f(t;\theta)\right)^2,
\end{equation}
and, hence, the partial derivative with respect to $x_0$ is given by:
\begin{equation}
    \frac{\partial \mathcal{L}}{\partial x_0} = \frac{1}{\sigma^2} (\rho + f_{x_0}(1) - \rho f_{x_0}(0)) \left(x_1 - \rho x_0 - (f(1;\theta) - \rho f(0;\theta)\right) + \frac{1}{\sigma^2} \sum_{t=2}^T \nu(t) (f_{x_0}(t;\theta) - \rho f_{x_0}(t;\theta)).
\end{equation}
The off-diagonal elements of the FIM are, thus, given by:
\begin{equation}
    \begin{split}
    \mathbb{E}\left[-\frac{\partial^2 \mathcal{L}}{\partial x_0 \partial \theta_i}\right] &= \frac{1}{\sigma^2}(\rho + f_{x_0}(1) - \rho f_{x_0}(0))(f_{\theta_i}(1) - \rho f_{\theta_i}(0)) +  \frac{1}{\sigma^2} \sum_{t=2}^T (f_{\theta_i}(t;\theta) - \rho f_{\theta_i}(t;\theta))(f_{x_0}(t;\theta) - \rho f_{x_0}(t;\theta))\\
    &= \frac{\rho}{\sigma^2}(f_{\theta_i}(1) - \rho f_{\theta_i}(0)) + \frac{1}{\sigma^2}\sum_{t=1}^T (f_{\theta_i}(t;\theta) - \rho f_{\theta_i}(t;\theta))(f_{x_0}(t;\theta) - \rho f_{x_0}(t;\theta))
    \end{split}
\end{equation}
Since these terms are not generally zero, uncertainty in $x_0$ affects uncertainty in model parameters, $\theta_i$. The diagonal term in the FIM corresponding to $x_0$ is given by:
\begin{equation}
    \begin{split}
    \mathbb{E}\left[-\frac{\partial^2 \mathcal{L}}{\partial x_0^2}\right] &= \frac{1}{\sigma^2}(\rho + f_{x_0}(1) - \rho f_{x_0}(0))^2 +  \frac{1}{\sigma^2} \sum_{t=2}^T (f_{x_0}(t;\theta) - \rho f_{x_0}(t;\theta))^2\\
    &= \frac{1}{\sigma^2}(\rho^2 + 2\rho (f_{x_0}(1) - \rho f_{x_0}(0))) +  \frac{1}{\sigma^2} \sum_{t=1}^T (f_{x_0}(t;\theta) - \rho f_{x_0}(t;\theta))^2
    \end{split}
\end{equation}

\section{MA(1) processes}\label{sec:appendix_ma}
Using the lag operator $L a_t = a_{t-1}$, eq. \eqref{eq:constant_ma1} can be rearranged as,
\begin{equation}\label{eq:koyck}
\nu(t) = \frac{1}{1+\phi L} (x(t) - \mu).
\end{equation}
Since $\nu(t)\stackrel{\text{IID}}{\sim}\mathcal{N}(0,\sigma)$ and using eq. \eqref{eq:koyck}, we can write the log-likelihood as,
\begin{equation}\label{eq:ma1_loglikelihood}
\mathcal{L} = -\frac{T}{2} \text{log }2\pi - \frac{T}{2} \text{log } \sigma^2 - \frac{1}{2\sigma^2}\sum_{t=2}^{T} \frac{1}{(1+\phi L)^2} \left(x(t) - \mu \right)^2,
\end{equation}
which yields the diagonal element of the information matrix,
\begin{equation}\label{eq:information_ma1}
\begin{aligned}
\mathcal{I}_{\mu,\mu} &= -\mathbb{E}(\frac{\partial^2 \mathcal{L}}{\partial \mu^2})\\
&= \frac{T}{\sigma^2(1+\phi)^2}.
\end{aligned}
\end{equation}
Eq. \eqref{eq:information_ma1} implies an asymptotic variance of the maximum likelihood estimator of $\mu$ given by,
\begin{equation}
\text{var}(\hat{\mu}) = \frac{\sigma^2(1+\phi)^2}{T}.
\end{equation}
We next derive the variance of estimates for $\mu$ under the false model -- that is, assuming that errors are independent Gaussians. Eq. \eqref{eq:false_error} gives the variance of this estimator in terms of the estimated variance $\sigma'^2$. In an infinite sample size, this variance converges to the true variance of the MA(1) process, meaning that the variance of the misspecified model is,
\begin{equation}
\text{var}(\tilde{\mu}) = \frac{\sigma^2(1+\phi^2)}{T}.
\end{equation}
The VIR is, hence, given by,
\begin{equation}\label{eq:ma1_vir}
\text{VIR}(\mu) = 1 + \frac{2\phi}{1+\phi^2}.
\end{equation}
To illustrate the validity of eq. \eqref{eq:ma1_vir}, we perform a series of synthetic data simulations using a constant mean model (eq. \eqref{eq:simple_system}) but with MA(1) errors (eq. \eqref{eq:ma1}). In these, we generate time series of length 1000, setting $\sigma=0.1$ and $\mu=10$ and assuming a range of $\phi$ values; we supply the actual $\phi$ values to the model (i.e. we do not fit this parameter) in each case and set priors on $\mu\sim U(-100, 100)$ and $\sigma\sim U(0,100)$. At each $\phi$ value, we perform 10 replicates; in each case, fitting both a model assuming MA(1) errors and another assuming IID errors. Using these two model fits, we then estimate the VIR by taking the ratio of the MA(1) model variance to that from the IID model. Models were coded up in Stan using a ``Generative model conditional likelihood'' approach similar to that described in \S\ref{sec:model_fitting}, and the code to reproduce these is in the Github repo. To fit the models, we use Stan's NUTS algorithm, with 4 Markov chains: using 2000 iterations per chain for the MA(1) model and 1000 for the IID model. In both cases, half of the iterations were discarded as warm-up. 

\begin{figure}[ht]
	\centerline{\includegraphics[width=1\textwidth]{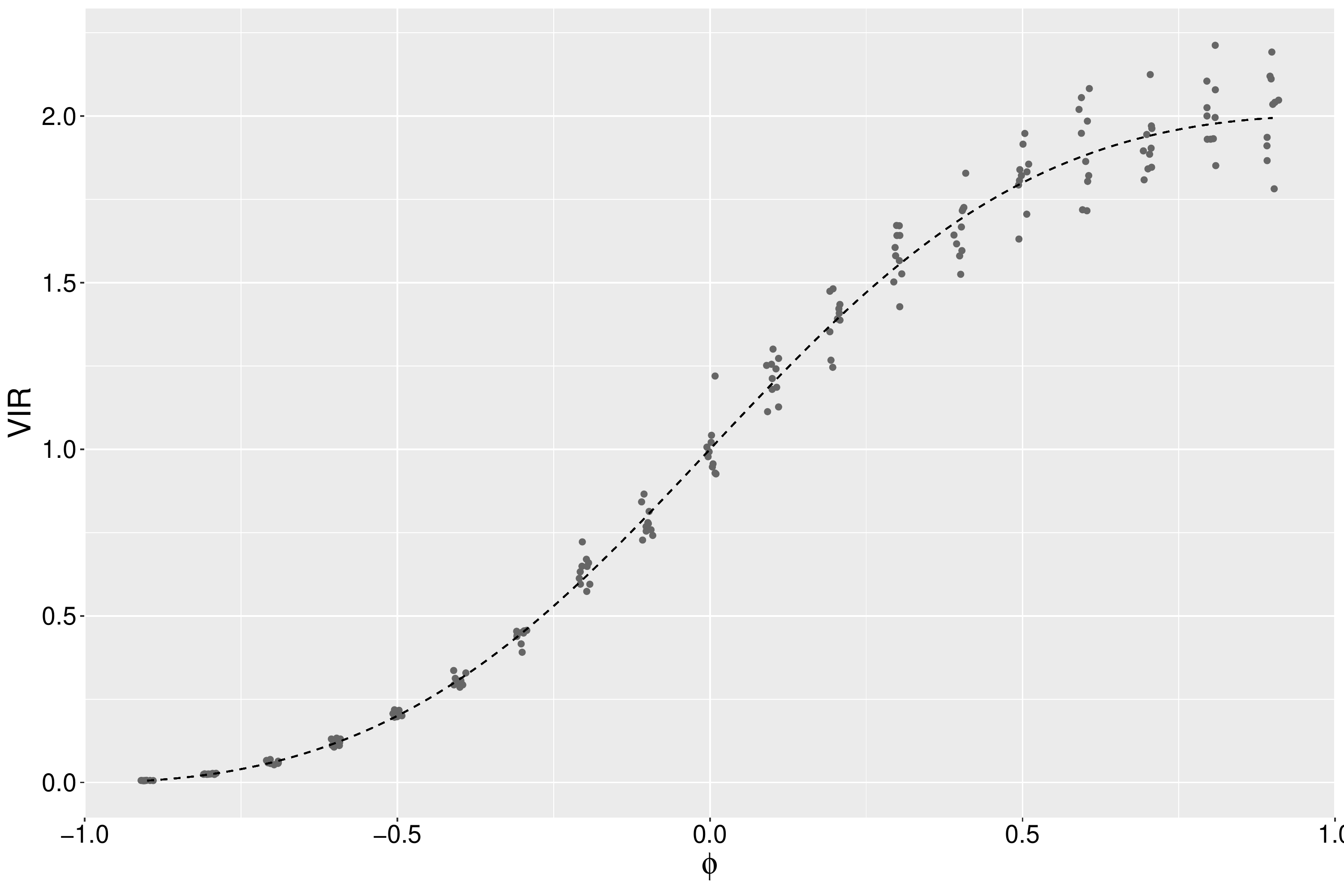}}
	\caption{\textbf{MA(1) model: VIRs.} The horizontal axis indicates the values of $\phi$ used to generate synthetic datasets using MA(1) errors as described in \S\ref{sec:appendix_ma}. The vertical axis shows the VIRs. Points show the estimated VIR from each replicate and the dashed line shows the function given by eq. \eqref{eq:ma1_vir}. Note, horizontal jitter has been added to points.}
	\label{fig:mar1_virs}
\end{figure}

Fig. \ref{fig:mar1_virs} shows the results of these simulations. Here, points indicate the estimated VIR from each replicate and the dashed line shows the analytic VIR given in eq. \eqref{eq:ma1_vir}. Across the range of $\phi$ explored, the mean of the estimated VIRs were well described by eq. \eqref{eq:ma1_vir}. At higher values of $\phi$, the synthetic results had greater variation because of the higher variance in the simulated data at these values.

Finally, we briefly discuss eq. \eqref{eq:koyck} as we are aware readers may not be familiar with the lag operator. For most purposes, the lag operator can be treated as any other algebraic quantity. Eq. \eqref{eq:koyck} can then be expanded out assuming that $|\phi|<1$ using the result for the sum of all terms of an infinite geometric series:
\begin{equation}
\begin{aligned}
(1 + \phi L)^{-1} (x(t) - \mu) &= (1 - \phi L + \phi^2 L^2 - \phi^3 L^3 + ...)(x(t) - \mu)\\
&=  (x(t) - \mu) - \phi (x(t-1) - \mu) + \phi^2(x(t-2) - \mu)\\
&\;\;\;- \phi^3(x(t-3) - \mu).
\end{aligned}
\end{equation}
This means that eq. \eqref{eq:ma1_loglikelihood} cannot simply be minimised by setting $\phi\rightarrow\infty$, as might be thought on first appearances.

\section{ARMA(1,1) processes}\label{sec:appendix_arma11}
In this section, we describe simulations we performed to check the validity of eq. \eqref{eq:arma11_vir}. In these, we generate time series of length 1000, setting $\sigma=0.1$ and $\mu=10$ and assuming a range of $\phi$ and $\rho$ values; we supply the actual $\phi$ and $\rho$ values to the model (i.e. we do not fit these parameters) in each case and set priors on $\mu\sim U(-100, 100)$ and $\sigma\sim U(0,100)$. At each $(\phi,\rho)$ value, we perform 10 replicates; in each case, fitting both a model assuming ARMA(1,1) errors and another assuming IID errors. Using these two model fits, we then estimate the VIR by taking the ratio of the ARMA(1,1) model variance to that from the IID model. Models were coded up in Stan using the ``Generative model conditional likelihood'' approach described in \S\ref{sec:model_fitting}, and the code to reproduce these is in the Github repo. To fit the models, we use Stan's NUTS algorithm, with 4 Markov chains: using 2000 iterations per chain for the ARMA(1,1) model and 1000 for the IID model. In both cases, half of the iterations were discarded as warm-up.

\begin{figure}[ht]
	\centerline{\includegraphics[width=1\textwidth]{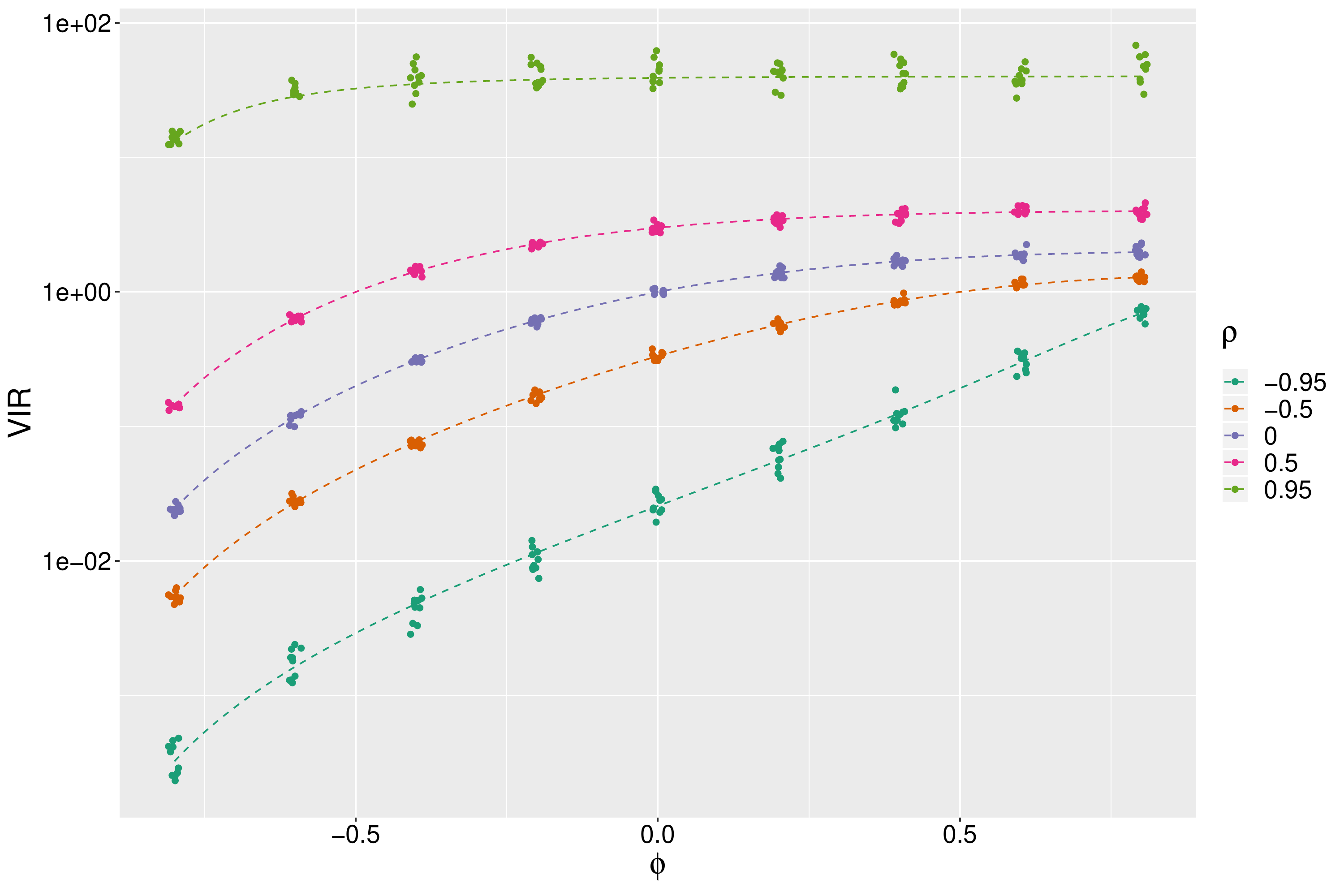}}
	\caption{\textbf{ARMA(1,1) model: VIRs.} The horizontal axis indicates the values of $\phi$ used to generate synthetic datasets using ARMA(1,1) errors as described in \S\ref{sec:appendix_arma11}; colours show the values of $\rho$ used (points and line colouring corresponds). The vertical axis shows the VIRs. Points show the estimated VIR from each replicate and the coloured lines show the function given by eq. \eqref{eq:arma11_vir}. Note, horizontal jitter has been added to points and that the vertical axis is on the log-scale.}
	\label{fig:arma11_virs}
\end{figure}

Fig. \ref{fig:arma11_virs} shows the results of these simulations. Here, points indicate the estimated VIR from each replicate and dashed lines shows the analytic VIR given by eq. \eqref{eq:arma11_vir}: the horizontal axis indicates $\phi$ values; colours indicate $\rho$ values. Across the range of parameters investigated, the theoretical VIRs are a good match to the empirical VIRs.

\section{Kalman filter likelihood}\label{sec:kalman}
In this section, we provide an alternative approach to fitting the ARMA(1,1) model described in \S\ref{sec:model_fitting}. The Kalman filter provides an alternative likelihood which does not require assuming that the first two terms $\nu(t)$ terms in eq. \eqref{eq:arma11} are zero (as in eq. \eqref{eq:conditional_loglikelihood}). The following borrows heavily from \cite{zivot2006state}, which we found to be a useful reference for Kalman filters. The key to using Kalman filters is to first get the model into a ``state-space'' form, and we do this using $\epsilon(t) = x(t) - f(t;\theta)$. Then, we define $\boldsymbol{\alpha}(t) = (\epsilon(t), \phi \nu(t))'$, which we then use to restate eq. \eqref{eq:arma11},

\begin{equation}\label{eq:transition}
\boldsymbol{\alpha}(t) = \boldsymbol{T}\boldsymbol{\alpha}(t-1) + \boldsymbol{R} \nu(t),
\end{equation}

\noindent where $\boldsymbol{T} = \begin{bmatrix} \rho & 1\\ 0 & 0 \end{bmatrix}$ and $\boldsymbol{R} = \begin{bmatrix}1 & 0\\0 & \phi \end{bmatrix}$. Eq. \eqref{eq:transition} is known as the ``transition equation'' for the system. The ``measurement equation'' is given by:
\begin{equation}\label{eq:measurement}
x(t) = \boldsymbol{Z} \boldsymbol{\alpha}(t) + f(t;\theta),
\end{equation}
where $\boldsymbol{Z}=(1, 0)$. To help with the derivation of the Kalman filter, we define the optimal predictor of $\boldsymbol{\alpha}(t)$ as ${\boldsymbol{a}(t) = \mathbb{E}[\boldsymbol{\alpha}(t)|I(t)]}$, where $I(t)$ denotes the information available at time $t$; we also define the mean square error (MSE) matrix describing uncertainty in predictions,

\begin{equation}
\boldsymbol{P}(t) = \mathbb{E}[(\boldsymbol{\alpha}(t) - \boldsymbol{a}(t))(\boldsymbol{\alpha}(t) - \boldsymbol{a}(t))'|I(t)].
\end{equation}

Given these, we can determine the optimal predictors of $\boldsymbol{\alpha}(t)$ and the associated MSE matrix given information at time $t-1$,
\begin{equation}
\begin{aligned}
\boldsymbol{a}(t|t-1) &= \mathbb{E}[\boldsymbol{\alpha}(t)|I(t-1)] = \boldsymbol{T}\boldsymbol{a}(t-1)\\
\boldsymbol{P}(t|t-1) &=  \mathbb{E}[(\boldsymbol{\alpha}(t) - \boldsymbol{a}(t|t-1))(\boldsymbol{\alpha}(t) - \boldsymbol{a}(t|t-1))'|I(t-1)].
\end{aligned}
\end{equation}
The expression,
\begin{equation}
\begin{aligned}
\boldsymbol{\alpha}(t) - \boldsymbol{a}(t|t-1) &= \boldsymbol{T}\boldsymbol{\alpha}(t-1) + \boldsymbol{R}\nu(t) - \boldsymbol{T}\boldsymbol{a}(t-1)\\
&=  \boldsymbol{R}\nu(t) + \boldsymbol{T}(\boldsymbol{\alpha}(t-1)-\boldsymbol{a}(t-1)),
\end{aligned}
\end{equation}
means that $\boldsymbol{P}(t|t-1) = \sigma^2\boldsymbol{R}\boldsymbol{R}' + \boldsymbol{T}\boldsymbol{P}(t-1)\boldsymbol{T}'$.

Now, we describe the ``updating'' equations, that allow determination of $\boldsymbol{a}(t)$ and $\boldsymbol{P}(t)$ from  $\boldsymbol{a}(t|t-1)$ and  $\boldsymbol{P}(t|t-1)$. To motivate the eventual expressions, we start by considering obtaining the first update at time $t=2$: from eq. \eqref{eq:transition}, the state vector at this time is,
\begin{equation}
\boldsymbol{\alpha}(2) = \boldsymbol{T}\boldsymbol{\alpha}(1) + \boldsymbol{R} \nu(2).
\end{equation}
Assuming $\boldsymbol{\alpha}(1)\sim\mathcal{N}(\boldsymbol{a}(1), \boldsymbol{P}(1))$, and because $\nu(t)$ is also normally distributed,
\begin{equation}
\boldsymbol{\alpha}(2) \sim \mathcal{N}(\boldsymbol{a}(2|1), \boldsymbol{P}(2|1)),
\end{equation}
where $\boldsymbol{a}(2|1) = \boldsymbol{T}\boldsymbol{a}(1)$ and $\boldsymbol{P}(2|1)=\boldsymbol{T} \boldsymbol{P}(1) \boldsymbol{T}' + \sigma^2\boldsymbol{R}\boldsymbol{R}'$. The measurement equation for this period is dictated by eq. \eqref{eq:measurement}, which implies that $x(2)$ is also normally distributed. We next define the optimal step ahead prediction for $x(t|t-1) = \boldsymbol{Z}\boldsymbol{a}(t|t-1)+ f(t;\theta)$, which we use to determine the joint distribution of $(\alpha(2), x(2))$,
\begin{equation}\label{eq:joint}
\begin{aligned}
\boldsymbol{\alpha}(2)  &= \boldsymbol{a}(2|1) + (\boldsymbol{\alpha}(2)-\boldsymbol{a}(2|1))\\
x(2)  &= x(2|1) + (x(2)-x(2|1))\\
&= \boldsymbol{Z}\boldsymbol{a}(2|1)+ f(2;\theta) + \boldsymbol{Z}(\boldsymbol{\alpha}(2) - \boldsymbol{a}(2|1)).
\end{aligned}
\end{equation}
Using the pair of expressions in \eqref{eq:joint}, we can determine the covariance,
\begin{equation}
\begin{aligned}
\text{Cov}(\boldsymbol{\alpha}(2), x(2)) &= \mathbb{E}[(\boldsymbol{\alpha}(2) - \boldsymbol{a}(2|1))(x(2) - \boldsymbol{Z}\boldsymbol{a}(2|1) - f(2;\theta))']\\
&= \mathbb{E}[(\boldsymbol{\alpha}(2)-\boldsymbol{a}(2|1))(\boldsymbol{Z}(\boldsymbol{\alpha}(2) - \boldsymbol{a}(2|1)))']\\
&= \mathbb{E}[(\boldsymbol{\alpha}(2)-\boldsymbol{a}(2|1))(\boldsymbol{\alpha}(2) - \boldsymbol{a}(2|1))']\boldsymbol{Z}'\\
&= \boldsymbol{P}(2|1)\boldsymbol{Z}'.
\end{aligned}
\end{equation}
Using this result, we can write down the joint distribution,

\begin{equation}
\begin{pmatrix}
\boldsymbol{\alpha}(2) \\
x(2)
\end{pmatrix}\sim \mathcal{N}\left(\begin{pmatrix}
\boldsymbol{a}(2|1) \\
\boldsymbol{Z}\boldsymbol{a}(2|1) + f(2;\theta)
\end{pmatrix},\begin{pmatrix}
\boldsymbol{P}(2|1) &  \boldsymbol{P}(2|1)\boldsymbol{Z}'\\
\boldsymbol{Z}\boldsymbol{P}(2|1) & \boldsymbol{Z}\boldsymbol{P}(2|1)\boldsymbol{Z}'
\end{pmatrix}\right).
\end{equation}
We then use standard results for conditional distributions of multivariate normals (see, for example, \cite{wiki2020normal}) to obtain $\boldsymbol{\alpha}(2)|x(2)\sim \mathcal{N}(\boldsymbol{a}(2), \boldsymbol{P}(2))$, where,
\begin{align}\label{eq:mean_update}
\boldsymbol{a}(2) &= \mathbb{E}(\boldsymbol{\alpha}(2)|x(2)) = \boldsymbol{a}(2|1) + \boldsymbol{P}(2|1) \boldsymbol{Z}'F(2)^{-1} v(2)\\
\label{eq:var_update}
\boldsymbol{P}(2) &=  \boldsymbol{P}(2|1) -  \boldsymbol{P}(2|1) \boldsymbol{Z}'F(2)^{-1} \boldsymbol{Z}\boldsymbol{P}(2|1).
\end{align}
In eqs. \eqref{eq:mean_update}\&\eqref{eq:var_update}, $v(2) = x(2) - x(2|1)$ represents the prediction error on the observable, and ${F(2) =  \mathbb{E}(v(2)v(2)') = \boldsymbol{Z} \boldsymbol{P}(2|1)\boldsymbol{Z}'}$, is the mean square error in this prediction.

Eqs. \eqref{eq:mean_update}\&\eqref{eq:var_update} generalise to future time periods, yielding the optimal filtering equations:
\begin{align}\label{eq:mean_update1}
\boldsymbol{a}(t) &= \boldsymbol{a}(t|t-1) + \boldsymbol{P}(t|t-1) \boldsymbol{Z}'F(t)^{-1} v(t)\\
\label{eq:var_update1}
\boldsymbol{P}(t) &=  \boldsymbol{P}(t|t-1) -  \boldsymbol{P}(t|t-1) \boldsymbol{Z}'F(t)^{-1} \boldsymbol{Z}\boldsymbol{P}(t|t-1).
\end{align}

The corresponding optimal predictor value of $x(t)$ given information at time $t-1$ is then,
\begin{equation}
x(t|t-1) = \boldsymbol{Z}\boldsymbol{a}(t|t-1) + f(t;\theta),
\end{equation}
which has predictive errors given by,
\begin{equation}
v(t) = x(t) - x(t|t-1) = x(t) - \boldsymbol{Z}\boldsymbol{a}(t|t-1) - f(t;\theta) = \boldsymbol{Z} (\boldsymbol{\alpha}(t) - \boldsymbol{a}(t|t-1)).
\end{equation}
The prediction variance is then given by,
\begin{equation}
F(t) = \mathbb{E}[v(t)v(t)'] =  \boldsymbol{Z} \boldsymbol{P}(t|t-1)\boldsymbol{Z}'.
\end{equation}
The predictive errors, $v(t) \sim \mathcal{N}(0, \sqrt{F(t)})$, and the log-likelihood can then be written as,
\begin{equation}
\mathcal{L} = -\frac{T}{2} \text{log}\; 2\pi - \frac{1}{2}\sum_{t=1}^{T}\text{log}\; F(t) - \frac{1}{2} \sum_{t=1}^{T} v(t)' F(t)^{-1} v(t).
\end{equation}

\section{Results}

\begin{table}[ht]
	\centering
	\begin{tabular}{|c|c|c|} 
		\hline
		Model & Parameter & Prior \\ \hline
		Both & $r$ & $T{\text -}\mathcal{N}(1.0, 1.0)$ \\
		& $\kappa$ & $\mathcal{N}(50.0,20.0)$ \\
		& $\bar{X}(0)$ & $T{\text -}\mathcal{N}(1.0, 0.5)$ \\ \hline
		IID & $\sigma$ & $T{\text -}\mathcal{N}(1.0/\sqrt{1 - \rho^2}, 1.0)$ \\
		\hline
	    AR(1) & $\sigma$ & $T{\text -}\mathcal{N}(1.0, 1.0)$ \\
		& $\rho$ & $\mathcal{N}(0.0, 0.5)$ \\ \hline
	\end{tabular}
	\caption{\textbf{Logistic model: prior parameters.} Here, ``$T{\text -}\mathcal{N}$'' denotes a normal distribution truncated to have support only over positive values. The $r$ and $\kappa$ had an upper bound of 100 specified, $\bar{X}(0)$ had an upper bound of 10, and $\rho$ had one at 0.99 to prevent the Markov chains in Stan initialising at unrealistic parameter values. Note, $\rho$ enters the IID model prior for $\sigma$ to ensure that we assume the same marginal prior variance for the errors in both models.}  \label{table:logistic_priors}
\end{table}

\begin{table}[ht]
	\centering
	\begin{tabular}{|c|c|c|} 
		\hline
		Model & Parameter & Prior \\ \hline
		Both & $g_{Kr}$ & $\text{log}{\text -}\mathcal{N}(10.5, 1.0)$ \\
		& $p_1$ & $\text{log}{\text -}\mathcal{N}(-2.5, 3.0)$ \\
		& $p_2$ & $\text{log}{\text -}\mathcal{N}(4.5, 1.0)$ \\
		& $p_3$ & $\text{log}{\text -}\mathcal{N}(-3.5, 1.5)$ \\
		& $p_4$ & $\text{log}{\text -}\mathcal{N}(4.0, 0.5)$ \\
		& $p_5$ & $\text{log}{\text -}\mathcal{N}(4.5, 0.5)$ \\
		& $p_6$ & $\text{log}{\text -}\mathcal{N}(3.0, 1.5)$ \\
		& $p_7$ & $\text{log}{\text -}\mathcal{N}(2.0, 0.5)$ \\
		& $p_8$ & $\text{log}{\text -}\mathcal{N}(3.5, 0.5)$ \\
		\hline
		IID & $\sigma$ & $\Gamma(2.5, 0.05)$ \\
		\hline
		AR(1) & $\sigma$ & $\Gamma(2.5, 0.05)$ \\
		& $\rho$ & $\text{beta}(4, 2)$\\ \hline
	\end{tabular}
	\caption{\textbf{hERG model: prior parameters.} Here, ``$\text{log}{\text -}\mathcal{N}()$'' indicates log-normal priors.}  \label{table:herg_priors}
\end{table}

\begin{table}[ht]
	\centering
	\begin{tabular}{|l|l|l|l|}
		\hline
		\multicolumn{1}{|c|}{\textbf{Symbol}} & \multicolumn{1}{c|}{\textbf{Description}} & \multicolumn{1}{c|}{\textbf{Units}} & \multicolumn{1}{c|}{\textbf{Value}} \\ \hline
		$c_{inf}$ & Far-field concentration of A & $\text{mol cm}{-3}$ & 1e-6 \\ \hline
		$D$ & Diffusion constant & $\text{cm}^2\text{ s}^{-1}$ & 7.2e-6 \\ \hline
		$F$ & Faraday constant & $\text{C mol}^{-1}$ & 96485.3328959 \\ \hline
		$R$ & Gas constant & $\text{J K}^{-1}\text{ mol}^{-1}$ & 96485.3328959 \\ \hline
		$S$ & Electrode area& $\text{cm}^{2}$ & 0.07 \\ \hline
		$T$ & Temperature & K & 297.0 \\ \hline
		$E_{start}$ & Input voltage start & V & 0.5 \\ \hline
		$E_{reverse}$ & Input voltage reverse & V & -0.1 \\ \hline
		$\Delta E$ & Input voltage amplitude & V & 0.03 \\ \hline
		$v$ & Scan rate & $\text{V s}^{-1}$ & 0.08941 \\ \hline
	\end{tabular}
	\caption{Electrochemistry model: (fixed) experimental parameters.}
	\label{tab:electrochem_experimental_parameters}
\end{table}

\begin{table}[ht]
	\centering
	\begin{tabular}{|l|l|l|l|l|}
		\hline
		\multicolumn{1}{|c|}{\textbf{Symbol}} & \multicolumn{1}{c|}{\textbf{Description}} &
		\multicolumn{1}{c|}{\textbf{Units}} & \multicolumn{1}{c|}{\textbf{Lower Bound}} &
		\multicolumn{1}{c|}{\textbf{Upper Bound}} \\ \hline
		$E_0$ & Reversible potential & V & 0.0 & 1.0 \\ \hline
		$k_0$ & Reaction rate & $\text{s}^{-1}$ & 0.0 & 0.4 \\ \hline
		$\alpha$ & Symmetry factor &  & 0.2 & 0.7 \\ \hline
		$C_{dl}$ & Capacitance & F & 1e-6 & 100e-6 \\ \hline
		$R_u$ & Uncompensated resistance & $\Omega$ & 0.0 & 1.0 \\ \hline
	\end{tabular}
	\caption{Electrochemistry model: priors.}
	\label{tab:electrochem_inference_parameters}
\end{table}

\subsection{Electrochemistry model}\label{sec:appendix_electrochemistry}
The field of voltammetry is concerned with interrogating and analysing redox reactions at electrodes. An input voltage signal is applied to an electrochemical cell comprising of the chemical species in solution around an electrode. This species reacts at the electrode surface, generating an output current signal. Inferring the parameters of the electrochemical model gives insight into the properties of the redox reaction taking place. We consider an electrochemistry model of a quasi-reversible redox reaction occurring in solution, given
by

\begin{align} \label{reaction}
A + e^- \cee{&<=>[E_0,k_0,\alpha]} B,
\end{align}

\noindent where species $A$ and $B$ are in solution, and the parameters of interest, $E_0$,
$k_0$, and $\alpha$, are the reversible formal potential, standard heterogeneous charge
transfer rate constant at $E_0$ and the charge transfer coefficient, respectively. 

As the diffusion of $A$ and $B$ are assumed to be equal, we only need to solve for the
concentration of a single species $c_A$, which diffuses across a 1-dimensional domain
using Fick's second law:

\begin{align}
\frac{\partial c_A}{\partial t} &= D \frac{\partial^2 c_A}{\partial x^2}, \label{diffeqtn}
\end{align}

\noindent where $x$ is distance from the electrode surface and $t$ is time. The
initial and boundary conditions are

\begin{align}
c_A(x,0) &= c_{\infty} \nonumber \\
c_A &\rightarrow c_{\infty},  \quad \text{as} \quad x \rightarrow \infty,\quad t>0. \label{icsandbcs}
\end{align}

At the electrode surface, $x=0$, for $t>0$, we have the conservation and flux conditions

\begin{align}
D \frac{\partial c_A}{\partial x}=\frac{I_f}{FS}, \label{fluxatelectrode}
\end{align}

\noindent along with the Butler-Volmer condition

\begin{align}
D \frac{\partial c_A}{\partial x} = \text{ }&k_0
\left[c_A\exp\left(-\alpha\frac{F}{RT}(E_{\mbox{\tiny
		eff}}(t)-E_0)\right) 
- (c_{\infty}-c_A)\exp\left( (1-\alpha) \frac{F}{RT}(E_{\mbox{\tiny
		eff}}(t)-E_0)\right)
\right]. \label{BVeqtn}
\end{align}

Here, $I_f$ is the faradaic current, $S$ is the electrode area, and
$E_{\mbox{\tiny eff}}(t)$ is the {\em effective} applied potential (defined
below).

We complete the model by defining $E_{\mbox{\tiny{app}}}(t)$ to be the applied
potential, which is given by the addition of a linear ramp and a sinusoidal term

\begin{align}
E_{\mbox{\tiny{app}}}(t) =  E_{\mbox{\tiny{start}}}  \left\lbrace
\begin{array}{ll}
+ v t + \Delta E \sin {(\omega t)}, \qquad 0 \le t \le t_{\mbox{\tiny
		reverse}}, \\
-  v t + 2 v t_{\mbox{\tiny reverse}} + \Delta E \sin {(\omega t)},
\qquad  t_{\mbox{\tiny reverse}}\le t \le 2t_{\mbox{\tiny reverse}},
\end{array}
\right.
\label{Eac}
\end{align}

\noindent where $v$ is the sweep rate,  $E_{\mbox{\tiny{start}}}$ is the initial
potential, $t_{\mbox{\tiny reverse}}$ is the time of switching from the forward to the
reverse sweep in cyclic voltammetry, $\omega$ is the radial frequency and $\Delta E$ is
the amplitude of the sine wave. The {\em effective} applied potential can now be defined
as

\begin{align}
E_{\mbox{\tiny{eff}}}(t) = E_{\mbox{\tiny{app}}}(t) - E_{\mbox{\tiny{drop}}} = E_{\mbox{\tiny{app}}}(t) - I_{\mbox{\tiny{tot}}} R_u, \nonumber
\end{align}

\noindent where $E_{\mbox{\tiny{drop}}}$ models the effect of uncompensated
resistance,
$R_u$. $I_{\mbox{\tiny{tot}}}$ is the total (measured) current, and combines the
faradaic current and the background capacitive current, $I_c$, which can be
modelled as

\begin{align}
I_c &= C_{dl}\frac{dE_{\mbox{\tiny{eff}}}}{dt}, \label{Ic}
\end{align}

where $C_{dl}$ is the double layer capacitance (assumed constant in this work), and then

\begin{align}
I_{\mbox{\tiny{tot}}} = I_f + I_c. \label{Itot}
\end{align}

The details of non-dimensionalising and solving the system of equations given by eqs. \eqref{diffeqtn}-\eqref{Itot} are given in the supplementary information of \cite{sher2004resistance}. In summary, the PDE in eq. \eqref{diffeqtn} is discretised using a finite difference method on an exponentially expanding grid. The time-stepping is performed using a backwards Euler method, combined with a Newton-Raphson method to solve for the non-linear boundary condition at the electrode surface.

The details of the experimental setup are given in \cite{morris2015theoretical}. Voltammetry was performed in a standard 3-electrode cell, using a glassy carbon macrodisk. The number of experimental time points measured was 525,000. This was evenly subsampled by a factor of 21 in order to reduce the number of time points to 25,000 and to reduce the computational cost of model fitting.

\end{document}